\documentclass[12pt]{article}
\usepackage{a4}
\usepackage{amssymb}
\usepackage{cite}
\newcommand{\be}{\begin{equation}}
\newcommand{\ee}{\end{equation}}
\newcommand{\bea}{\begin{eqnarray}}
\newcommand{\eea}{\end{eqnarray}}

\usepackage{hyperref}
\begin{document}
\thispagestyle{empty}
\def\thefootnote{\fnsymbol{footnote}}
\begin{center}\Large
On classical and semiclassical properties of \\
the Liouville theory with defects. \\
\vskip 2em
May 2015
\end{center}\vskip 0.2cm
\begin{center}
Hasmik Poghosyan$^{1}$
\footnote{hasmikpoghos@gmail.com}
 and Gor Sarkissian$^{1,2}$\footnote{ gor.sarkissian@ysu.am}
\end{center}
\vskip 0.2cm

\begin{center}
$^1$ Yerevan Physics Institute, \\
Alikhanian Br. 2, 0036\, Yerevan\\
Armenia
\end{center}
\begin{center}
$^2$Department of Physics, \ Yerevan State University,\\
Alex Manoogian 1, 0025\, Yerevan\\
Armenia
\end{center}

\vskip 1.5em
\begin{abstract} \noindent
The Lagrangian of the Liouville theory with topological defects is analyzed in detail
and general solution of the corresponding  defect equations of motion is found.
We study the heavy and light semiclassical limits of the defect two-point function
found before via the bootstrap program.
We show  that the heavy asymptotic limit is given by the exponential of the Liouville
action with defects, evaluated on the solutions with two singular points.
We demonstrate that the light asymptotic limit
is given by the finite dimensional path integral over solutions of the defect equations of motion
with a vanishing energy-momentum tensor.
\end{abstract}
\newpage
\tableofcontents
\newpage
\section{Introduction}

Defects in two-dimensional conformal field theories can be realized as oriented lines, separating
different theories. We are interested in the special class of defects, for which the energy-momentum tensor
is continuous across the defect \cite{Petkova:2000ip}. Denoting  the left- and right- moving energy-momentum tensors of the two theories by $T^{(1)}$, $T^{(2)}$, and $\bar{T}^{(1)}$, $\bar{T}^{(2)}$,
this condition takes the form:
\be
T^{(1)}=T^{(2)}\, ,\hspace{1cm} \bar{T}^{(1)}=\bar{T}^{(2)} \label{TopDef11}\, .
\ee
Inserting a defect  in the path integral is equivalent in the operator language to the insertion of an operator $D$ which maps the Hilbert space of CFT $1$ to that of CFT $2$.
  Condition (\ref{TopDef11}) can be considered
as  implying that the corresponding operator $D$ commutes with the Virasoro modes:
\be\label{topcom}
D L^{(1)}_m=L^{(2)}_m D\hspace{1cm}{\rm and}\hspace{1cm} D \bar{L}^{(1)}_m=\bar{L}^{(2)}_m D\, .
\ee
During the last few years topological defects in the Liouville and Toda field theories attracted some attention due to their
relation to the Wilson lines in the AGT correspondence \cite{Alday:2009fs,Drukker:2009id,Petkova:2009pe,Drukker:2010jp}
\footnote{In fact in references \cite{Alday:2009fs,Drukker:2009id} the Verlinde loop operators are discussed,
but they coincide with topological defects for the Cardy case \cite{Fuchs:2010hk}.}.
Defects in the Liouville field theory have been constructed in \cite{Sarkissian:2009aa,Sarkissian:2011tr}.
In these papers defects were constructed as operators on the Hilbert space of Liouville theory.
To obtain these operators, two-point functions in the presence of defects were calculated using
the conformal bootstrap program for defects, developed in  \cite{Sarkissian:2009aa,Petkova:2001ag}.
It was shown in \cite{Sarkissian:2009aa} that there are two families of defects: discrete,  corresponding to the degenerate fields
and labeled by a pair of positive integers $m$ and $n$, with eigenvalues
\be\label{discret1}
{\cal D}_{m,n}(\alpha)={\sin(\pi m b^{-1}(2\alpha- Q))\sin(\pi nb (2\alpha-Q))\over \sin\pi b^{-1}(2\alpha-Q)\sin\pi b(2\alpha-Q)}\, ,
\ee

 and continuous, labeled by one continuous parameter $s$ with eigenvalues
 \be\label{cont1}
{\cal D}_{s}(\alpha)={\cosh(2\pi s(2\alpha-Q))\over 2\sin\pi b^{-1}(2\alpha-Q)\sin\pi b(2\alpha-Q)}\, .
\ee
We denoted here by $Q=b+{1\over b}$ the background charge, and $\alpha$ labels primaries of
Liouville theory.
The defects of the discrete family have a one-dimensional world-volume, and in particular
the identity defect ${\cal D}_{1,1}$ belongs to the discrete family.
The defects of the continuous family have a two-dimensional world-volume.
The details can be found in appendix D.

Recently also integrable defects were studied (see e.g.
\cite{Bowcock:2003dr,Bowcock:2004my,Corrigan:2009vm,Aguirre:2015xsa,Aguirre:2013zfa,Caudrelier:2008zz,Caudrelier:2014gsa,Robertson:2014rma,Avan:2011sr}
).

The Lagrangian for the continuous family of two-dimensional topological defects was suggested in \cite{Aguirre:2013zfa}.
It is demonstrated in \cite{Aguirre:2013zfa} that topological defects are so called type-II defects, proposed in \cite{Corrigan:2009vm}, allowing
additional degrees of freedom associated with the defect itself.
It is  also shown in \cite{Aguirre:2013zfa}, that requiring the additional degrees of freedom to be represented by a holomorphic field, leads to
the topological defects.

The aim of this work is to study correspondence between the continuous family of defects (\ref{cont1})
and the one-parametric family of Lagrangians with defect proposed in \cite{Aguirre:2013zfa}.

First we find general solution of the defect equations of motion coming from
the Lagrangian proposed in \cite{Aguirre:2013zfa}.

To link two-point functions in the presence of defects to the Lagrangian with defects we use
two strategies: heavy and light asymptotic  semiclassical limits
\cite{Seiberg:1990eb,Zamolodchikov:1995aa,Fateev:2007ab,Fateev:2010za,Hadasz:2006vs,Menotti:2006gc,Menotti:2006tc}.
In the light asymptotic limit we set $\alpha=\eta_l b$ and keep $\eta_l$ fixed for $b\to 0$, whereas
in the heavy asymptotic limit we take $\alpha={\eta_h\over b}$ and hold $\eta_h$ fixed again for $b\to 0$.

These semiclassical limits were used in \cite{Zamolodchikov:1995aa,Fateev:2007ab} to relate
 the quantum three-point functions in the Liouville and Toda theories
with the corresponding classical  actions.
The heavy asymptotic limit plays an important role in the quantum uniformization program \cite{Ginsparg:1993is}.
In papers \cite{Fateev:2010za,Hadasz:2006vs,Menotti:2006gc,Menotti:2006tc}
these techniques were generalized to the boundary Liouville and Toda theories.
Both limits have recently proved to be very useful also to test AGT \cite{Mironov:2009qn,Fateev:2011qa,Hama:2013ama,Piatek:2013ifa} and AdS/CFT correspondences \cite{Hijano:2015qja,Alkalaev:2015wia,Fitzpatrick:2015zha}.

The heavy and light asymptotic limits were reconsidered  in \cite{Harlow:2011ny} also for complex solutions of the analytically continued Liuoville theory.

Here we develop both procedures of the semiclassical limits to the Liouville theory with defects and
find perfect agreement between the classical and bootstrap results. In particular we establish
connection between the parameter $\kappa$ entering in the Lagrangian with defect and parameter $s$
labeling the defect operator (\ref{cont1}):
\be
\kappa=\cosh(2\pi s b)
\ee
where it is understood that $s\to \infty $ and $b\to 0$ in a way that keeps $\sigma=sb$  fixed.

We show that in the light asymptotic limit the defect two-point functions can be obtained via the path integral
over solutions of the defect equations of motion with  vanishing energy-momentum tensor in the large
$\sigma$ limit.

We demonstrate that in the heavy asymptotic limit defect two-point functions are given by
the sum of exponentials of the action with defects evaluated on solutions with two singular points of the defect equations of motion.
To understand better the semiclassical origin of the denominator in (\ref{cont1}) in the heavy asymptotic limit, we consider analytic continuation of
 $\eta$ to the complex region in the spirit of \cite{Harlow:2011ny}.
 We find a discrete family of solutions with two singular points, labelled by two integer numbers $N_1$ and $N_2$.
 But to fit to semiclassical limit of the defect two-point function and to have convergent series we should sum over the saddle points with nonnegative $N_1$ and $N_2$ for ${\rm Im}(2\eta-1)>0$, and with nonpositive $N_1$ and $N_2$ if
${\rm Im}(2\eta-1)<0$. This is an example of the Stokes phenomena \cite{Witten:2010cx,Harlow:2011ny,Marino:2012zq,Pasquetti:2009jg}.

The paper is organized in the following way.

In section 2 we analyze classical Liouville theory with defects. In subsection 2.1 we review the general solution
of the Liouville equation. In subsection 2.2 we present general solution of the defect equations of motion.
In subsection 2.3 we present the Lagrangian of the  product of the Liouville theories on half-plane with the
boundary condition specified by a permutation brane.
In section 3 we review defects and permutation branes in quantum Liouville theory.
In section 4 we review the heavy and light asymptotic semiclassical limits.
In section 5 we calculate the defect two-point function in the light asymptotic limit.
In section 6 we calculate   the defect two-point function in the heavy asymptotic limit.
In a series of appendices we describe some useful technical results.

\section{Classical Liouville theory with defects}
\subsection{Review of Liouville solution}
Let us recall some facts on classical Liouville theory.

The action of the Liouville theory is

\be\label{azione}
S={1\over 2\pi i}\int\left(\partial\phi\bar{\partial} \phi+\mu \pi e^{2b\phi}\right)d^2 z\, .
\ee
Here we use a complex coordinate $z=\tau+i\sigma$, and $d^2z\equiv dz\wedge d\bar{z}$ is the volume form.

The field $\phi(z,\bar{z})$ satisfies the classical Liouville equation of motion
\be\label{leom}
\partial\bar{\partial}\phi=\pi\mu b e^{2b\phi}\, .
\ee
The general solution to  (\ref{leom}), also derived below, was given by Liouville in terms
of two arbitrary functions $A(z)$ and $B(\bar{z})$ \cite{liouville}
\be\label{liusol}
\phi={1\over 2b}\log\left({1\over \pi\mu b^2}{\partial A(z)\bar{\partial}B(\bar{z})
\over (A(z)+B(\bar{z}))^2}\right)\, .
\ee
The solution (\ref{liusol}) is invariant if one transforms $A$ and $B$
simultaneously by the  following constant M\"{o}bius transformations:
\be\label{mobsim}
A\to{\zeta A+\beta\over \gamma A+\delta},\hspace{0.5cm}
B\to{\zeta B-\beta\over -\gamma B+\delta},\hspace{0.5cm}
\zeta\delta-\beta\gamma=1\, .
\ee
Classical expressions for left and right components of the energy-momentum tensor are
\be\label{tl}
T=-(\partial\phi)^2+b^{-1}\partial^2\phi\, ,
\ee
\be\label{tr}
\bar{T}=-(\bar{\partial}\phi)^2+b^{-1}\bar{\partial}^2\phi\, .
\ee
Substituting (\ref{liusol}) in (\ref{tl}) and (\ref{tr})
we get, that the components of the energy-momentum tensor are
 given  by the Schwarzian derivatives of $A(z)$ and $B(\bar{z})$:
\be
T=\{A; z\}={1\over 2b^2}\left[{A'''\over A'}-{3\over 2}{(A'')^2\over (A')^2}\right]\, ,
\ee
\be
\bar{T}=\{B; \bar{z}\}={1\over 2b^2}\left[{B'''\over B'}-{3\over 2}{(B'')^2\over (B')^2}\right]\, .
\ee
The Schwarzian derivative is invariant under arbitrary constant
 M\"{o}bius transformation:
\be
\left\{{\zeta F+\beta\over \gamma F+\delta}; z\right\}=\{F;z\}
,\hspace{0.5cm}
\zeta\delta-\beta\gamma=1\, .
\ee

Solutions of the Liouville equation (\ref{leom}) can be described also
via linear combination of some holomorphic and anti-holomorphic functions.
Let us introduce the function $V=e^{-b\phi}$.
One can write the Liouville equation  (\ref{leom}) as an equation for $V$
\be\label{vip}
V\partial\bar{\partial}V-\partial V\bar{\partial}V=-\pi\mu b^2\, .
\ee
Also the left and right components of the energy-momentum tensor (\ref{tl}) and  (\ref{tr}) can be written via $V$
\be\label{vta1}
\partial^2V=-b^2VT\, ,
\ee
\be\label{vtb1}
\bar{\partial}^2V=-b^2V\bar{T}\, .
\ee
It is straightforward to check that the general solution of  eq. (\ref{vip}) is  given
by linear combination of two holomorphic $a_i(z)$, $i=1,2$, and two anti-holomorphic functions $b_i(\bar{z})$, $i=1,2$:
\be\label{ab12}
V=\sqrt{\pi\mu b^2}\Bigg(a_1(z)b_1(\bar{z})-a_2(z)b_2(\bar{z})\Bigg)\, ,
\ee
satisfying the condition
\be\label{prwro}
(a_1a'_2-a'_1a_2)(b_1b'_2-b'_1b_2)=1\, .
\ee
Usually the fields $a_i(z)$ and $b_i(\bar{z})$, $i=1,2$ are normalized to have  the unit
Wronskian:
\be\label{a1a2}
a_1a'_2-a'_1a_2=1
\ee
and
\be\label{b1b2}
b_1b'_2-b'_1b_2=1\, .
\ee
It is easy to see that the left and right components of the energy-momentum tensor
can be expressed via $a_i$ and $b_i$ in the very simple form:
\be\label{tbaa}
T=-{1\over b^2}{\partial^2 a_1\over a_1}=-{1\over b^2}{\partial^2 a_2\over a_2}
\ee
and
\be\label{tabb}
\bar{T}=-{1\over b^2}{\bar{\partial}^2 b_1\over b_1}=-{1\over b^2}{\bar{\partial}^2 b_2\over b_2}\, .
\ee
The solutions (\ref{liusol}) and (\ref{ab12}) can be related in the following way.
One can solve the unit Wronskian conditions (\ref{a1a2}) and (\ref{b1b2})
via a holomorphic $A(z)$ and an anti-holomorphic function $B(\bar{z})$
\be\label{AAA}
a_1={1\over \sqrt{\partial A}}\hspace{0.5cm}{\rm and}\hspace{0.5cm} a_2={A\over \sqrt{\partial A}}
\ee
and
\be\label{BBB}
b_1={B\over \sqrt{\bar{\partial} B}}\hspace{0.5cm}{\rm and}\hspace{0.5cm} b_2=-{1\over \sqrt{\bar{\partial} B}}\, .
\ee
Inserting (\ref{AAA})  and (\ref{BBB}) in (\ref{ab12}) we get (\ref{liusol}).
Note that the M\"{o}bius transformations of $A$ and $B$ (\ref{mobsim}) become linear $SL(2,\mathbb{C})$ transformations of $a_i$ and $b_i$:

\bea\label{delga}
&&\tilde{a}_1=\delta a_1+\gamma a_2\, ,\\ \nonumber
&&\tilde{a}_2=\beta a_1+\zeta a_2
\eea

and
\bea\label{albe}
&&\tilde{b}_1=\zeta b_1+\beta b_2\, ,\\ \nonumber
&&\tilde{b}_2=\gamma b_1+\delta b_2\, .
\eea

It is straightforward to check that indeed (\ref{ab12}) is invariant under  (\ref{delga})
and (\ref{albe}), and both of them keep the unit Wronskian condition.

One can also check, that both components of the energy-momentum tensor (\ref{tbaa}) and (\ref{tabb}) are
invariant under these transformations as well.

We finish this  section with a remark which will be important in the parts of this work dealing with the light asymptotic limit.
There we will consider an analytic continuation $\mu\to -\mu$. At this point it is convenient to write the solution (\ref{ab12})
  as:
\be\label{ab132}
V=\sqrt{-\pi\mu b^2}\Bigg(a_1(z)b_1(\bar{z})+a_2(z)b_2(\bar{z})\Bigg)\, .
\ee
It is easy to check that (\ref{ab132}) also solves the Liouville equation, given that $a_i$ and $b_i$, $i=1,2$
obey the condition (\ref{prwro}).

\subsection{Lagrangian of the Liouville theory with defect}
Recently in  \cite{Aguirre:2013zfa} the action of the Liouville theory
with topological defects was suggested:
\bea\label{topdef}
&&S^{\rm top-def}={1\over 2\pi i}\int_{\Sigma_1}\left(\partial\phi_1\bar{\partial} \phi_1+\mu \pi e^{2b\phi_1}\right)d^2 z+
{1\over 2\pi i}\int_{\Sigma_2}\left(\partial\phi_2\bar{\partial} \phi_2+\mu \pi e^{2b\phi_2}\right)d^2 z\\ \nonumber
&&+ \int_{\partial \Sigma_1}\left[-{1\over 2\pi}\phi_2\partial_{\tau}\phi_1+
{1\over 2\pi}\Lambda\partial_{\tau}(\phi_1-\phi_2)+{\mu\over 2}
e^{(\phi_1+\phi_2-\Lambda)b}-{1\over \pi b^2}
e^{\Lambda b}\left(\cosh(\phi_1-\phi_2)b-\kappa\right)\right]{d\tau\over i}\, .
\eea
Here $\Sigma_1$ is the upper half-plane $\sigma={\rm Im} z\geq 0$ and
$\Sigma_2$ is the lower  half-plane $\sigma={\rm Im} z\leq 0$.
The defect is located along their common boundary,
which is the real axis $\sigma=0$ parametrized by $\tau={\rm Re} z$.
Note that $\Lambda(\tau)$ here is an additional
field associated with the defect itself.
The action (\ref{topdef}) yields the following defect equations of motion at $\sigma=0$:
\be\label{eom1}
{1\over 2\pi}(\partial-\bar{\partial})\phi_1+
{1\over 2\pi}\partial_{\tau}\phi_2-
{1\over 2\pi}\partial_{\tau}\Lambda+
{\mu b\over 2}e^{(\phi_1+\phi_2-\Lambda)b}
-{1\over \pi b}e^{\Lambda b}\sinh (\phi_1-\phi_2)b=0\, ,
\ee

\be\label{eom2}
-{1\over 2\pi}(\partial-\bar{\partial})\phi_2-
{1\over 2\pi}\partial_{\tau}\phi_1+
{1\over 2\pi}\partial_{\tau}\Lambda+
{\mu b\over 2}e^{(\phi_1+\phi_2-\Lambda)b}
+{1\over \pi b}e^{\Lambda b}\sinh (\phi_1-\phi_2)b=0\, ,
\ee
\be\label{eom3}
{1\over 2\pi}\partial_{\tau}(\phi_1-\phi_2)-{\mu b\over 2}
e^{(\phi_1+\phi_2-\Lambda)b}-{1\over \pi b}
e^{\Lambda b}\left(\cosh(\phi_1-\phi_2)b-\kappa\right)=0\, .
\ee
The last equation is derived from variation of $\Lambda$.

Using that $\partial_{\tau}=\partial+\bar{\partial}$
and  forming various linear combinations of  equations (\ref{eom1})-(\ref{eom3})
we can bring them
to the form:
\be\label{def3}
\bar{\partial}(\phi_1-\phi_2)=
\pi\mu b e^{b(\phi_1+\phi_2)}e^{-\Lambda b}\, ,
\ee
\be\label{kapik}
\partial(\phi_1-\phi_2)={2\over b}e^{\Lambda b}
\left(\cosh(\phi_1-\phi_2)b-\kappa\right)\, .
\ee
\be\label{def11}
\partial(\phi_1+\phi_2)-\partial_{\tau}\Lambda=
{2\over b}e^{\Lambda b}\sinh(b(\phi_1-\phi_2))\, .
\ee
It is shown in  \cite{Aguirre:2013zfa} that requiring the
defect equations of motion to hold for every $\sigma$ brings
additionally to the condition, that $\Lambda$ is a restriction to the real axis of a holomorphic
field
\be\label{hollam}
\bar{\partial}\Lambda=0\, .
\ee
This condition allows to rewrite (\ref{def11}) in the form
\be\label{def111}
\partial(\phi_1+\phi_2-\Lambda)=
{2\over b}e^{\Lambda b}\sinh(b(\phi_1-\phi_2))\, .
\ee
It is checked in \cite{Aguirre:2013zfa}
that the system of the defect equations of motion (\ref{def3})-(\ref{def111})
guarantees that both components of the energy-momentum tensor
are continuous across the defects and therefore
describes
topological defects:
 \be\label{tl1}
-(\partial\phi_1)^2+b^{-1}\partial^2\phi_1=
-(\partial\phi_2)^2+b^{-1}\partial^2\phi_2\, ,
\ee
\be\label{tr2}
-(\bar{\partial}\phi_1)^2+b^{-1}\bar{\partial}^2\phi_1=
-(\bar{\partial}\phi_2)^2+b^{-1}\bar{\partial}^2\phi_2\, .
\ee
Another interesting consequence of the defect equations of motion,
found in \cite{Aguirre:2013zfa}, is the existence together with the holomorphic field
$\Lambda(z)$ of an anti-holomorphic  field $\Xi$:
\be
\partial\Xi=0\, ,
\ee
where
\be
\Xi=e^{-b(\phi_1+\phi_2)}e^{b\Lambda}(\cosh b(\phi_1-\phi_2)
-\kappa)\, .
\ee
or alternatively
\be\label{xi}
\Xi={b\over 2}e^{-b(\phi_1+\phi_2)}\partial(\phi_1-\phi_2)\, .
\ee

Now we will present the general solution for defect equations of motion (\ref{def3})-(\ref{def111}).

We will follow essentially the same strategy which was used
in  \cite{Gervais:1981gs} for analyzing the boundary Liouville problem.
On the one hand since the defect is topological both components of the
energy-momentum tensor are equal being computed in terms
of $\phi_1$ or $\phi_2$. On the other hand
each component of the energy-momentum tensor
is given by the Schwarzian derivative, which is invariant under
the M\"{o}bius transformation.
This naturally leads to the following solution:
\be
\phi_1={1\over 2b}\log\left({1\over \pi\mu b^2}{\partial A\bar{\partial}B
\over (A+B)^2}\right)\, ,
\ee

\be
\phi_2={1\over 2b}\log\left({1\over \pi\mu b^2}{\partial C\bar{\partial}D
\over (C+D)^2}\right)\, ,
\ee
where
\be
C={\zeta A+\beta\over \gamma A+\delta}\hspace{0.5cm}
{\rm and} \hspace{0.5cm} D={\zeta' B+\beta'\over \gamma' B+\delta'}\, .
\ee
Remembering  that  $\phi_2$ is invariant under
the simultaneous   M\"{o}bius transformation (\ref{mobsim}) of $C$ and $D$, we can
set $B=D$. Therefore without loosing generality we can look for
a solution in the form:
\be\label{phi1}
\phi_1={1\over 2b}\log\left({1\over \pi\mu b^2}{\partial A\bar{\partial}B
\over (A+B)^2}\right)\, ,
\ee

\be\label{phi2}
\phi_2={1\over 2b}\log\left({1\over \pi\mu b^2}{\partial C\bar{\partial}B
\over (C+B)^2}\right)\, ,
\ee
where
\be\label{cphik}
C={\zeta A+\beta\over \gamma A+\delta}\, .
\ee
Substituting (\ref{phi1}) and (\ref{phi2}) in (\ref{def3})
we find that it is satisfied with
\be\label{lambdik}
e^{-\Lambda b}={A-C\over \sqrt{\partial A\partial C}}\, .
\ee
Since $A$ and $C$ are holomorphic functions, $\Lambda$
is holomorphic as well, as it is stated in (\ref{hollam}).

It is straightforward to check that (\ref{def111}) is satisfied
as well with $\phi_1$, $\phi_2$ and $\Lambda$
given by
 (\ref{phi1}), (\ref{phi2}) and (\ref{lambdik})
respectively.
And finally inserting (\ref{phi1}), (\ref{phi2}) and (\ref{lambdik})
in (\ref{kapik})  we see that it is also fulfilled with
\be\label{aldet}
\kappa={\zeta+\delta\over 2}\, .
\ee
Inserting (\ref{phi1}), (\ref{phi2}) in (\ref{xi}) one can check that
\be
\Xi={\pi\mu b^2\over 2}{\gamma B^2+B(\zeta-\delta)-\beta
\over \bar{\partial}B}\, .
\ee
Remembering that $B$ is anti-holomorphic we see
that $\Xi$ is anti-holomorphic as well.

We can also write the solution of the defect equations of motion  using solution of the Liouville equation in the form (\ref{ab12}).
Recalling that the M\"{o}bius transformations of the functions $A$ and $B$ become linear $SL(2,\mathbb{C})$
transformations of the functions $a_i$ and $b_i$, which leave the components of the energy-momentum tensor (\ref{tbaa}) and (\ref{tabb})
invariant,
we can write the solution  (\ref{phi1})-(\ref{cphik}) in the form:
\be
e^{-b\phi_1}=\sqrt{\pi\mu b^2}\Bigg(a_1(z)b_1(\bar{z})-a_2(z)b_2(\bar{z})\Bigg)\, ,
\ee
\be
e^{-b\phi_2}=\sqrt{\pi\mu b^2}\Bigg(c_1(z)b_1(\bar{z})-c_2(z)b_2(\bar{z})\Bigg)\, ,
\ee
where denoting $\vec{a}=(a_1,a_2)$, $\vec{c}=(c_1,c_2)$, and $D=\left(\begin{array}{cc}
\delta&\gamma\\
\beta&\zeta \end{array}\right)$, one has
\be
\vec{c}=D\vec{a}
\ee
and
\be\label{trd}
2\kappa={\rm Tr}\ D\, .
\ee

At this point we would like to make the following remark.
Let us consider the identity defect. It has $A=C$, and $\kappa=1$.
Setting $A=C$ in (\ref{lambdik}) we obtain $e^{-\Lambda b}=0$.
This result can be derived also directly setting $\phi_1=\phi_2$ in (\ref{def3}).
Therefore the identity defect does not belong to the family of defects described
by the action (\ref{topdef}) and can be derived from them only in the limit
$\Lambda\to \infty$.
This can be understood recalling from appendix D that defects described by
(\ref{topdef}) have a two-dimensional world-volume in a sense that  the values
of $\phi_1(\tau)$ and $\phi_2(\tau)$ at an arbitrary point $\tau$ on the defect line
are not constrained and the point $(\phi_1(\tau),\phi_2(\tau))$ can take
values in the whole plane $\mathbb{R}^2$. Contrary to this, the identity defect
has a one-dimensional world-volume, since the point $(\phi_1(\tau),\phi_2(\tau))$
takes values on one-dimensional diagonal $\phi_1=\phi_2$.

\subsection{Lagrangian of the Liouville theory with permutation branes}
We can also  construct a folded version of the action
(\ref{topdef}) describing product of Liouville theories on a half-plane
with boundary condition given by permutation branes:
\bea
&&S^{\rm perm-brane}=
{1\over 2\pi i}\int_{\Sigma}\left(\partial\phi_1\bar{\partial} \phi_1+\mu \pi e^{2b\phi_1}+
\partial\phi_2\bar{\partial} \phi_2+\mu \pi e^{2b\phi_2}\right)d^2 z
\\ \nonumber
&&+ \int_{\partial \Sigma}\left[-{1\over 2\pi}\phi_2\partial_{\tau}\phi_1+
{1\over 2\pi}\Lambda\partial_{\tau}(\phi_1-\phi_2)-{\mu\over 2}
e^{(\phi_1+\phi_2-\Lambda)b}+{1\over \pi b^2}
e^{\Lambda b}\left(\cosh(\phi_1-\phi_2)b-\kappa\right)\right]{d\tau\over i}\, .
\eea
 $\Sigma$ denotes here the upper half-plane $\sigma\geq 0$, and $\tau$ parameterizes the boundary located at
 $\sigma=0$.
This action gives rise to the boundary equations

\be\label{eom1p}
{1\over 2\pi}(\partial-\bar{\partial})\phi_1+
{1\over 2\pi}\partial_{\tau}\phi_2-
{1\over 2\pi}\partial_{\tau}\Lambda-
{\mu b\over 2}e^{(\phi_1+\phi_2-\Lambda)b}
+{1\over \pi b}e^{\Lambda b}\sinh (\phi_1-\phi_2)b=0\, ,
\ee

\be\label{eom2p}
{1\over 2\pi}(\partial-\bar{\partial})\phi_2-
{1\over 2\pi}\partial_{\tau}\phi_1+
{1\over 2\pi}\partial_{\tau}\Lambda-
{\mu b\over 2}e^{(\phi_1+\phi_2-\Lambda)b}
-{1\over \pi b}e^{\Lambda b}\sinh (\phi_1-\phi_2)b=0\, .
\ee
\be\label{eom3p}
{1\over 2\pi}\partial_{\tau}(\phi_1-\phi_2)+{\mu b\over 2}
e^{(\phi_1+\phi_2-\Lambda)b}+{1\over \pi b}
e^{\Lambda b}\left(\cosh(\phi_1-\phi_2)b-\kappa\right)=0\, .
\ee

Again using that  $\partial_{\tau}=\partial+\bar{\partial}$ and forming
various linear combinations, one can bring the system (\ref{eom1p})-(\ref{eom3p})
to the form
 \be\label{perma}
\partial\phi_2-\bar{\partial}\phi_1=
\pi\mu b e^{b(\phi_1+\phi_2)}e^{-\Lambda b}\, ,
\ee
\be\label{pursh}
\partial\phi_1-\bar{\partial}\phi_2=-{2\over b}e^{\Lambda b}
\left(\cosh(\phi_1-\phi_2)b-\kappa\right)\, ,
\ee
\be\label{perdef11}
\partial\phi_1+\bar{\partial}\phi_2-\partial_{\tau}\Lambda=
-{2\over b}e^{\Lambda b}\sinh(b(\phi_1-\phi_2))\, .
\ee
One can check that equations (\ref{perma})-(\ref{perdef11})  imply
the permutation brane conditions:
\bea
&&T^{(1)}=\bar{T}^{(2)}|_{\sigma=0}\, ,\\ \nonumber
&&\bar{T}^{(1)}=T^{(2)}|_{\sigma=0}
\eea
or using (\ref{tl}) and (\ref{tr})
\be\label{prtl1}
-(\partial\phi_1)^2+b^{-1}\partial^2\phi_1=
-(\bar{\partial}\phi_2)^2+b^{-1}\bar{\partial}^2\phi_2\, ,
\ee
\be\label{prtr2}
-(\bar{\partial}\phi_1)^2+b^{-1}\bar{\partial}^2\phi_1=
-(\partial\phi_2)^2+b^{-1}\partial^2\phi_2\, .
\ee

To solve equations (\ref{perma})-(\ref{perdef11}) we will use the same strategy as before, with the
only difference that now the M\"{o}bius transformation relates holomorphic and antiholomorphic functions:
\be\label{pa}
\phi_1={1\over 2b}\log\left({1\over \pi\mu b^2}{\partial A\bar{\partial}B
\over (A+B)^2}\right)\, ,
\ee

\be\label{pc}
\phi_2={1\over 2b}\log\left({1\over \pi\mu b^2}{\partial B\bar{\partial}C
\over (C+B)^2}\right)\, ,
\ee
and
\be\label{ca}
C={\zeta A+\beta\over \gamma A+\delta}\, .
\ee
The expressions (\ref{pa})-(\ref{ca}) solve  equation  (\ref{perma})
with the $\Lambda$ given by the relation
\be\label{pla}
e^{-\Lambda b}={C-A\over \sqrt{\partial A\bar{\partial} C}}\, .
\ee

It is straightforward to see that the expressions (\ref{pa})-(\ref{ca}) together with the $\Lambda$
given by (\ref{pla}) solve also eq. (\ref{perdef11}).

Finally inserting  $\phi_1$, $\phi_2$ and $\Lambda$ given by
(\ref{pa}), (\ref{pc}) and (\ref{pla}) respectively  in eq. (\ref{pursh})
we get that it is satisfied as well with the following $\kappa$

\be
\kappa={\zeta+\delta\over 2}\, .
\ee

\section{Permutation branes and defects in Quantum Liouville}

\subsection{Review of quantum Liouville}

Liouville  field theory is a conformal field theory enjoying the Virasoro algebra
\be
[L_m,L_n]=(m-n)L_{m+n}+{c_L\over 12}(n^3-n)\delta_{n,-m}\, ,
\ee
with the central charge
\be
c_L=1+6Q^2\, .
\ee

Primary fields $V_{\alpha}$ in this theory, which are associated with exponential fields
$e^{2\alpha \varphi}$, have conformal dimensions
\be
\Delta_{\alpha}=\alpha(Q-\alpha)\, .
\ee
The fields $V_{\alpha}$ and $V_{Q-\alpha}$ have the same conformal dimensions and represent
the same primary field, i.e. they are proportional to each other:
\be\label{valphr}
V_{\alpha}=S(\alpha)V_{Q-\alpha}\, ,
\ee
with the reflection function
\be\label{reflal}
S(\alpha)={\left(\pi\mu\gamma(b^2)\right)^{b^{-1}(Q-2\alpha)}\over b^2}{\Gamma(1-b(Q-2\alpha))\Gamma(-b^{-1}(Q-2\alpha))
\over \Gamma(b(Q-2\alpha))\Gamma(1+b^{-1}(Q-2\alpha))}\, .
\ee

Two-point functions of Liouville theory are given by the reflection function
(\ref{reflal}):
\be\label{twopf}
\langle V_{\alpha}(z_1,\bar{z}_1)V_{\alpha}(z_2,\bar{z}_2)\rangle={S(\alpha)\over (z_1-z_2)^{2\Delta_{\alpha}}
(\bar{z}_1-\bar{z}_2)^{2\Delta_{\alpha}}}\, .
\ee
Introducing ZZ function \cite{Zamolodchikov:2001ah}:
\be\label{zzfu}
W(\alpha)=-{2^{3/4}(\pi\mu\gamma(b^2))^{-{(Q-2\alpha)\over 2b}}\pi(Q-2\alpha)\over
\Gamma(1-b(Q-2\alpha))\Gamma(1-b^{-1}(Q-2\alpha))}\, ,
\ee
the two-point function can be compactly written as
\be\label{salpha}
S(\alpha)={W(Q-\alpha)\over W(\alpha)}\, .
\ee
Another useful property of ZZ function is
\be\label{wqaw}
W(Q-\alpha)W(\alpha)=-2\sqrt{2}\sin\pi b^{-1}(2\alpha-Q)\sin\pi b(2\alpha-Q)\, .
\ee
The spectrum of the Liouville theory has
the  form
\be\label{lsdi}
{\cal H}=\int_0^{\infty} dP \;R_{{Q\over 2}+iP}\otimes R_{{Q\over 2}+iP}\, ,
\ee
where $R_{\alpha}$ is the highest weight representation with respect to the Virasoro algebra.

\subsection{Permutation branes and defects in quantum Liouville}
Let us recall the form of continuous family of defects and permutation branes
in the Liouville field theory  computed in \cite{Sarkissian:2009aa,Sarkissian:2011tr} using
appropriate generalization of the Cardy-Lewellen equation \cite{Petkova:2001ag}.
The details can be found in appendix D.
Topological defects are intertwining operators $X$ commuting with the Virasoro generators
\be\label{lint}
[L_n,X]=[\bar{L}_n,X]=0\, .
\ee
Such operators have the form
\be
X=\int_{{Q\over 2}+i\mathbb{R}} d\alpha\,{\cal D}(\alpha)\,\mathbb{P}^{\alpha}\, ,
\ee
where $\mathbb{P}^{\alpha}$ are projectors on a subspace $R_{\alpha}\otimes R_{\alpha}$:
\be
\mathbb{P}^{\alpha}=\sum_{N,M}(|\alpha,N\rangle\otimes \overline{|\alpha,M\rangle})
(\langle \alpha,N|\otimes \overline{\langle\alpha,M|})\, .
\ee
Here $|\alpha,N\rangle$ and $\overline{|\alpha,M\rangle}$ are vectors of  orthonormal bases of the left and right copies of $R_{\alpha}$
respectively.
The eigenvalues ${\cal D}(\alpha)$ can be determined via the  two-point functions computed in the presence of a defect $X$
\be\label{defpoii}
 \langle V_{\alpha}(z_1,\bar{z}_1)XV_{\alpha}(z_2,\bar{z}_2)\rangle={{\cal D}(\alpha)S(\alpha)
\over (z_1-z_2)^{2\Delta_{\alpha}}(\bar{z}_1-\bar{z}_2)^{2\Delta_{\alpha}}}
\ee

It is shown in \cite{Sarkissian:2009aa}  that
\be\label{defpoi}
 \langle V_{\alpha}(z_1,\bar{z}_1)X_sV_{\alpha}(z_2,\bar{z}_2)\rangle=-{1\over W^2(\alpha)}
{2^{1/2}\cosh(2\pi s(2\alpha-Q))
\over (z_1-z_2)^{2\Delta_{\alpha}}(\bar{z}_1-\bar{z}_2)^{2\Delta_{\alpha}}}
\ee
and therefore for $D_s(\alpha)$ one can write using (\ref{salpha}) and (\ref{wqaw})
\be\label{dalsal}
{\cal D}_s(\alpha)=-{2^{1/2}\cosh(2\pi s(2\alpha-Q))\over S(\alpha)W^2(\alpha)}={\cosh(2\pi s(2\alpha-Q))\over 2\sin\pi b^{-1}(2\alpha-Q)\sin\pi b(2\alpha-Q)}\, .
 \ee
The parameter $s$ is a continuous parameter labeling the defect. The defects can be characterized also
by the value of the  two-point function of the  degenerate field $V_{-b/2}$ in the presence of a defect. It is a function $A(b)$ of $b$.
It is shown in \cite{Sarkissian:2009aa} that the parameter $s$ is related to the function $A(b)$ by the equation:
\be\label{relaa}
2\cosh 2\pi bs=A(b)\left({W(-b/2)\over W(0)}\right)^2\, .
\ee
The permutation branes boundary states $|B\rangle_{\cal P}$ on product $L_1\times L_2$ of two Liouville theories satisfy
the gluing condition \cite{Recknagel:2002qq}:
\bea\label{virgl}
&&(L_{n}^{(1)}-\bar{L}_{-n}^{(2)})|B\rangle_{\cal P}=0,\\ \nonumber
&&(L_{n}^{(2)}-\bar{L}_{-n}^{(1)})|B\rangle_{\cal P}=0.
\eea
Comparing the gluing conditions (\ref{virgl}) and (\ref{lint}) one can see that topological defects related
to permutation branes by folding trick, consisting of exchanging left and right components of the second copy,
and hence
these branes are characterized by the same two-point functions
(\ref{defpoi}) with $z_2$ and $\bar{z}_2$ exchanged
\be\label{perpoi}
\langle V_{\alpha}^{(1)}(z_1,\bar{z}_1)V_{\alpha}^{(2)}(z_2,\bar{z}_2)\rangle_{\cal P}=-{1\over W^2(\alpha)}
{2^{1/2}\cosh(2\pi s(2\alpha-Q))
\over (z_1-\bar{z}_2)^{2\Delta_{\alpha}}(\bar{z}_1-z_2)^{2\Delta_{\alpha}}}\, .
\ee




\section{Semiclassical limits}

\subsection{Heavy asymptotic limit}

Let us consider the action (\ref{azione})  for the rescaled variable $\varphi=2b\phi$
\be\label{azioneb}
S={1\over 8\pi ib^2}\int\left(\partial\varphi\bar{\partial} \varphi+4\lambda e^{\varphi}\right)d^2 z\, ,
\ee
where $\lambda=\pi\mu b^2$.

This form shows that $b^2$ plays in the Liouville theory the role of  the Planck constant, and
one can study semiclassical limit taking the limit $b\to 0$, in such a way that $\lambda$
is kept fixed.

Let us consider correlation functions in the path integral formalism:
\be\label{valpal}
\Big\langle V_{\alpha_1}(z_1,\bar{z}_1)\cdots V_{\alpha_n}(z_n,\bar{z}_n)\Big\rangle=
\int{\cal D}\varphi\; e^{-S}\;\prod_{i=1}^{n}\exp\left({\alpha_i\varphi(z_i,\bar{z}_i)\over b}\right)\, .
\ee
We would like to calculate this integral in the semiclassical limit $b\to 0$ using the method of steepest descent,
and we should decide how $\alpha_i$ scales with $b$. Since $S$ scales like $b^{-2}$, for operators to affect
the saddle point, we should take $\alpha_i=\eta_i/b$, with $\eta_i$ fixed.
The conformal weights $\Delta_{\alpha}=\eta(1-\eta)/b^2$ scale like $b^{-2}$ as well.
This is the heavy asymptotic limit.
Another choice of the operator scaling will be discussed in the next subsection.

We see from (\ref{valpal}) that in the semiclasscial limit the correlation function is given by $e^{-S_{\rm cl}}$
where, at least naively, in a sense which will be clarified below, $S_{\rm cl}$ is the  action
\be\label{naivact}
S={1\over 8\pi ib^2}\int\left(\partial\varphi\bar{\partial} \varphi+4\lambda e^{\varphi}\right)d^2 z
+\sum_{i=1}^{n}{\eta_i\over b^2}\varphi(z_i,\bar{z}_i)\, ,
\ee
evaluated on the solution of  its equation of motion:
\be\label{pusik}
\partial\bar{\partial}\varphi=2\lambda e^{\varphi}- 4\pi\sum_{i=1}^{n}\eta_i\delta^2(z-z_i)\, .
\ee
Assuming that in the vicinity of the insertion point $z_i$,  one can ignore the exponential term we get
that in the neighborhood of the  point $z_i$ $\varphi$ has the following behavior
\be\label{viscon}
\varphi(z,\bar{z})=-4\eta_i\log|z-z_i|+X_i \hspace{0.5cm}{\rm as } \hspace{0.5cm} z\to z_i\, .
\ee

One can insert this solution back into the equation of motion to check, if indeed
the exponential term is subleading. We find, that this happens when
\be\label{seibb}
{\rm Re}\,\eta_i<{1\over 2}\, .
\ee
This constraint is known as Seiberg bound \cite{Seiberg:1990eb}. It is the semiclassial version of the quantum condition
(\ref{valphr}) stating that $V_{\alpha}$ and $V_{Q-\alpha}$ represent the same quantum operator.
Either $\alpha$ or $Q-\alpha$ always obey the Seiberg bound.

Remembering that in the Liouville theory we have also a background charge at infinity,
conditions (\ref{viscon}) should be complemented by the behavior at the infinity:
 \be\label{infcon}
\varphi(z,\bar{z})=-2\log|z|^2 \hspace{0.5cm}{\rm as } \hspace{0.5cm} |z|\to \infty\, .
\ee
Since the energy-momentum tensor in the presence of  primary fields acquires a quadratic singularity,the functions $a_j$, $j=1,2$,  should solve the equation
\be
\partial^2a_j+b^2Ta_j=0\, ,
\ee
where
\be\label{tc12}
b^2T=\sum_{i=1}^{n}\left({\eta_i(1-\eta_i)\over (z-z_i)^2}+{c_i\over (z-z_i)}\right)
\ee
and $c_i$ are the so called accessory parameters.

If one tries naively to evaluate the action (\ref{naivact}) on a solution obeying
(\ref{viscon}), one finds that it diverges. Therefore we should consider a regularized action.
It was constructed in \cite{Zamolodchikov:1995aa}:
\bea\label{regact}
b^2S^{\rm reg}&=&{1\over 8\pi i}\int_{D-\cup_i d_i}\left(\partial\varphi\bar{\partial} \varphi+4\lambda e^{\varphi}\right)d^2 z
+{1\over 2\pi}\oint_{\partial D}\varphi d\theta+2\log R\\ \nonumber
&-&\sum_{i=1}^n\left({\eta_i\over 2\pi}\oint_{\partial d_i}\varphi d\theta_i+2\eta_i^2\log \epsilon_i\right)\, .
\eea
Here $D$ is a disc of radius $R$, $d_i$ is a disc of radius $\epsilon_i$ around $z_i$.
It was shown in \cite{Zamolodchikov:1995aa} that the action (\ref{regact}) satisfies the equation
\be\label{zamequ}
{\partial\over \partial \eta_i} b^2S^{\rm reg}=-X_i\, ,
\ee
where $X_i$ is defined by the boundary condition (\ref{viscon}).

The Polyakov conjecture proved in \cite{zograf}  states, that the action (\ref{regact}) obeys also  the relation:
\be\label{polyakus}
{\partial\over \partial z_i} b^2S^{\rm reg}=-c_i\, .
\ee
Let us write down a regularized version of the action with a defect.

First of all let us write it in  terms of $\lambda=\pi\mu b^2$, $\varphi_1=2b\phi_1$,  $\varphi_2=2b\phi_2$, and $\tilde{\Lambda}=2b\Lambda$:
\bea\label{topdefb}
&&b^2S^{\rm top-def}={1\over 8\pi i}\int_{\Sigma_1}\left(\partial\varphi_1\bar{\partial} \varphi_1+4\lambda e^{\varphi_1}\right)d^2 z+
{1\over 8\pi i}\int_{\Sigma_2}\left(\partial\varphi_2\bar{\partial} \varphi_2+4\lambda e^{\varphi_2}\right)d^2 z\\ \nonumber
&&+ \int_{\partial \Sigma_1}\left[-{1\over 8\pi}\varphi_2\partial_{\tau}\varphi_1+
{1\over 8\pi}\tilde{\Lambda}\partial_{\tau}(\varphi_1-\varphi_2)+{\lambda\over 2\pi}
e^{(\varphi_1+\varphi_2-\tilde{\Lambda})/2}-{1\over \pi }
e^{\tilde{\Lambda}/2}\left(\cosh\left({\varphi_1-\varphi_2\over 2}\right)-\kappa\right)\right]{d\tau\over i}\, .
\eea

Since we consider here only insertion of the bulk field, and do not consider insertion of the defect or boundary fields, the regularized
action takes the form:
\bea\label{topdefbb}
&&b^2S^{\rm top-def}={1\over 8\pi i}\int_{\Sigma_1^R-\cup_i d_i}\left(\partial\varphi_1\bar{\partial} \varphi_1+4\lambda e^{\varphi_1}\right)d^2 z\\ \nonumber
&-&\sum_{i=1}^n\left({\eta_i\over 2\pi}\oint_{\partial d_i}\varphi_1 d\theta_i+2\eta_i^2\log \epsilon_i\right)
+{1\over 2\pi}\int_{{s_R}_1}\varphi_1 d\theta+\log R\\ \nonumber
&+&{1\over 8\pi i}\int_{\Sigma_2^R-\cup_j d_j}\left(\partial\varphi_2\bar{\partial} \varphi_2+4\lambda e^{\varphi_2}\right)d^2 z\\ \nonumber
&-&\sum_{j=n+1}^{n+m}\left({\eta_j\over 2\pi}\oint_{\partial d_j}\varphi_2 d\theta_j+2\eta_j^2\log \epsilon_j\right)
+{1\over 2\pi}\int_{{s_R}_2}\varphi_2 d\theta+\log R\\ \nonumber
&+& \int_{-R}^R\left[-{1\over 8\pi}\varphi_2\partial_{\tau}\varphi_1+
{1\over 8\pi}\tilde{\Lambda}\partial_{\tau}(\varphi_1-\varphi_2)+{\lambda\over 2\pi}
e^{(\varphi_1+\varphi_2-\tilde{\Lambda})/2}-{1\over \pi }
e^{\tilde{\Lambda}/2}\left(\cosh\left({\varphi_1-\varphi_2\over 2}\right)-\kappa\right)\right]{d\tau\over i}\, ,
\eea
where $\Sigma_i^R$ is a half-disc of the radius $R$ and  ${s_R}_i$ is a semicircle of the radius $R$ in the half-plane $\Sigma_i$, $i=1,2$.

\subsection{Light asymptotic limit}
Another limit is the so called light asymptotic limit. Here we take
\be
\alpha=b\eta\, .
\ee
In this limit the operator insertions have no influence and the components of the energy-momentum
tensor are  (anti-) holomorphic and regular functions everywhere on sphere and hence vanish.
Eq.\@ (\ref{vta1}) and (\ref{vtb1}) imply that $V\equiv e^{-b\phi}$ should be at  most of first degree
in $z$ and $\bar{z}$, thus leading to the solutions
\footnote{It is shown in \cite{Fateev:2007ab} that to have solution in light limit
one needs to perform analytical continuation $\mu\to -\mu$.}
:
\be\label{vszz}
V(z,\bar{z};R)=\sqrt{-\lambda}(sz\bar{z}+tz+u\bar{z}+v)\, ,\hspace{1cm} R=\left(\begin{array}{cc}
s&t\\
u&v \end{array}\right)\, ,
\ee
where
\be
\label{det1}
{\rm det} R=sv-ut=1\, .
\ee

Therefore the path integral in the light limit becomes a finite-dimensional integral
over parameters $(s,t,u,v)$ which besides constraint (\ref{det1}) may
satisfy some additional constraints like reality and defect/boundary condition.
The reality of $V$ requires  the matrix $R$
to be Hermitian.
A way to parameterize the Hermitian matrices $R$ is
\be\label{x0x1}
R=\left(\begin{array}{cc}
X_0-X_1&X_2+iX_3\\
X_2-iX_3&X_0+X_1 \end{array}\right)\, ,
\ee
where the constraint $X_0^2-X_1^2-X_2^2-X_3^2=1$, makes clear that the moduli space of the real solutions of the Liouville equation (\ref{leom})
with the vanishing energy-momentum tensor is a three-dimensional hyperboloid
$H^{+}_3$.
Hence, for example in the bulk Liouville theory, the  correlation function in the light asymptotic limit takes the form
\be
\Big\langle V_{b\eta_1}(z_1,\bar{z}_1)\cdots V_{b\eta_n}(z_n,\bar{z}_n)\Big\rangle\to
e^{-S_l}\int_{H^{+}_3}dR\prod_{i=1}^{n}V^{-2\eta_i}(z_i,\bar{z}_i;R)\, ,
\ee
where $S_l$ is the value of the action on these solutions. The action $S_l$ is independent
on $R$, since the derivative of $S_l$ by any element of $R$ vanishes, thanks to
(\ref{vszz}) being solution of the equations of motion:
\be
{\partial S_l\over \partial R_{ij}}={\delta S_l\over \delta \phi}{\partial\phi\over\partial R_{ij}}=0
\ee
To avoid calculation of $S_l$ and some overall factors in the integration measure, it is more convenient, as suggested in \cite{Fateev:2010za}, to compute the ratio
\be
{\Big\langle V_{b\eta_1}(z_1,\bar{z}_1)\cdots V_{b\eta_n}(z_n,\bar{z}_n)\Big\rangle\over
\langle V_0(0)\rangle}\, .
\ee
Therefore defining
\be
\Big\langle V_{b\eta_1}(z_1,\bar{z}_1)\cdots V_{b\eta_n}(z_n,\bar{z}_n)\Big\rangle^{\rm light}\equiv
\int_{M}dR\prod_{i=1}^{n}V^{-2\eta_i}(z_i,\bar{z}_i;R)\, ,
\ee
where $M$ is the moduli space of solutions with a vanishing energy-momentum tensor satisfying
the corresponding boundary conditions in question,
we can write
\be
{\Big\langle V_{b\eta_1}(z_1,\bar{z}_1)\cdots V_{b\eta_n}(z_n,\bar{z}_n)\Big\rangle\over
\langle V_0(0)\rangle}\to
{\Big\langle V_{b\eta_1}(z_1,\bar{z}_1)\cdots V_{b\eta_n}(z_n,\bar{z}_n)\Big\rangle^{\rm light}\over
\langle V_0(0)\rangle^{\rm light}}\, .
\ee
The moduli space $M$ for the Liouville theory with a boundary was studied in \cite{Fateev:2010za}. It
was found that in the boundary Liouville problem $M$ is a subspace of $H_3^{+}$ with $X_3$ set to the boundary cosmological constant.
In the next section we will construct $M$ for the Liouville problem with defects.

\section{Defects in the light asymptotic limit}

Let us now specialize to the light asymptotic limit rules to the defects.
We should find solutions for $\phi_1$ and $\phi_2$ in the form (\ref{vszz}) satisfying the defect equations of motion.
We find it convenient to use in this section a new constant $\tilde{\lambda}\equiv -\lambda=-\pi\mu b^2$.
One can check that expressions
\bea\label{v1m1}
&&V_1(z,\bar{z}; R_1)\equiv e^{-b\phi_1}=\sqrt{\tilde{\lambda}}(s_1z\bar{z}+t_1z+u_1\bar{z}+v_1),\\ \nonumber
&&R_1=\left(\begin{array}{cc}
s_1&t_1\\
u_1&v_1 \end{array}\right),\ {\rm det}R_1=1\ , \, R_1^{\dagger}=R_1
\eea
and
\bea\label{v2m2}
&&V_2(z,\bar{z}; R_2)\equiv e^{-b\phi_2}=\sqrt{\tilde{\lambda}}(s_2z\bar{z}+t_2z+u_2\bar{z}+v_2),\\ \nonumber
&&R_2=\left(\begin{array}{cc}
s_2&t_2\\
u_2&v_2 \end{array}\right),\  {\rm det}R_2=1, \, R_2^{\dagger}=R_2
\eea
satisfy the defect equations of motion
(\ref{def3})-(\ref{def111})
with
\be\label{11t2}
2\kappa={\rm Tr}\left(R_2R_1^{-1}\right)= s_1v_2+s_2v_1-u_1t_2-u_2t_1
\ee
and
\be\label{s1t2}
e^{-b\Lambda}=z^2(s_1t_2-s_2t_1)+z(s_1v_2-s_2v_1+
u_1t_2-u_2t_1)+u_1v_2-u_2v_1\, .
\ee

Let us show that the relation (\ref{11t2}) results from the general formula (\ref{trd}).
Note that one can write the solution (\ref{v1m1})  in the general form (\ref{ab132})
\be
V_1(z,\bar{z}; R_1)=\sqrt{\tilde{\lambda}}(s_1z\bar{z}+t_1z+u_1\bar{z}+v_1)=
\sqrt{\tilde{\lambda}}[z(s_1\bar{z}+t_1)+(u_1\bar{z}+v_1)]\hspace{1cm}
\ee
with
\bea\label{absol}
&&a_1=z\, ,\hspace{2.2cm} a_2=1\, ,\\ \nonumber
&&b_1=s_1\bar{z}+t_1\, ,\hspace{1cm} b_2=u_1\bar{z}+v_1\, .
\eea
Remember that topological defects can be obtained in constructing $\phi_2$ by rotating the pair $a_1$, $a_2$
by a $SL(2,\mathbb{C})$ matrix $D=\left(\begin{array}{cc}
\zeta&\beta\\
\gamma&\delta \end{array}\right)$, namely taking
\bea\label{tildak}
&&\tilde{a}_1=\zeta z +\beta\, ,\\ \nonumber
&&\tilde{a}_2=\gamma z+\delta
\eea
and keeping the same $b_1$ and $b_2$ as in (\ref{absol}).
Using (\ref{tildak}) we get that $\phi_2$ is given by  $R_2=D R_1$.
Recalling that according to (\ref{trd}) $2\kappa={\rm Tr}\ D$
we arrive to  (\ref{11t2}).

We would like to mention also a folded version of the defect solution, obeying
the permutation brane  boundary conditions. One can see that
the expressions (\ref{v1m1}) and  (\ref{v2m2}) satisfy the permutation branes boundary conditions
(\ref{perma})-(\ref{perdef11})
 with
\be\label{kap12}
2\kappa={\rm Tr} (R_2^TR_1^{-1})=s_1v_2+s_2v_1-t_1t_2-u_1u_2
\ee
and
\be\label{lam21}
e^{-b\Lambda}=\tau^2(s_2t_1-s_1u_2)+\tau (s_2v_1-s_1v_2+
t_1t_2-u_1u_2)+t_2v_1-u_1v_2\, .
\ee

Note that equations (\ref{kap12}) and (\ref{lam21}) are in fact a folded version of  the corresponding
defect expressions (\ref{11t2}) and (\ref{s1t2}) derived by exchanging $u_2\leftrightarrow t_2$, as a result of
the $z_2\leftrightarrow \bar{z}_2$ exchange. The relation (\ref{kap12})  can be justified again using the general formalism
developed in section 2.3.

In the   parameterization (\ref{x0x1}) for  the Hermitian matrices $R_1$  and $R_2$
\be
R_1=\left(\begin{array}{cc}
X_0-X_1&X_2+iX_3\\
X_2-iX_3&X_0+X_1 \end{array}\right)\, ,\hspace{0.2cm}
R_2=\left(\begin{array}{cc}
Y_0-Y_1&Y_2+iY_3\\
Y_2-iY_3&Y_0+Y_1 \end{array}\right)\, ,
\ee
the defect parameter  (\ref{11t2})
is equal to the Minkowski inner product of the  vectors $X^{\mu}$ and $Y^{\mu}$
\be\label{kapka}
\kappa=X^{\mu}Y_{\mu}=X_0Y_0-X_1Y_1-X_2Y_2-X_3Y_3\, .
\ee
Using that $X_0,Y_0\geq 1$ and that $X^{\mu}$ and $Y^{\mu}$ both have the unit Minkowski norm, it is easy to show
that $X^{\mu}Y_{\mu}\geq 1 $, with equality when $X^{\mu}=Y^{\mu}$ \cite{Weinberg:1995mt}. It means that the real solutions of the defect equations of motion with
vanishing energy-momentum tensor exist only for $\kappa\geq 1$. The border at $\kappa=1$ is expected.
At this point $R_1=R_2$ and we have the identity defect  which has $e^{-b\Lambda}=0$, which
reflects that the identity defect does not belong to the family of two-dimensional defect described
by the action (\ref{topdef}).
It may happen that the semiclassical limit for other values of $\kappa$ can be obtained using complex
solutions of the defect equations of motion. Here we will consider only the values of $\kappa$ greater than $1$.

We are in a position to write the two-point correlation function in the presence of a defect:
\bea
&&\langle V_{\alpha}(z_1,\bar{z}_1)XV_{\alpha}(z_2,\bar{z}_2)\rangle^{\rm light} =\\ \nonumber
&&\int_{H_3^{+}\times  H_3^{+}}dR_1dR_2
\delta\Big({\rm Tr}\left(R_2R_1^{-1}\right)-2\kappa\Big)
V^{-2\eta}_1(z_1,\bar{z}_1; R_1)V^{-2\eta}_2(z_2,\bar{z}_2; R_2)\, .
\eea
Here $dR_i$, $i=1,2$ denotes integration measure on the 3D hyperboloid $H_3^{+}$.
This expression allows to  establish conformal invariance of a defect  two-point function.
Let us perform the transformation
\be\label{rrr}
R_1\to LR_1L^{\dagger}\hspace{1cm} {\rm and} \hspace{1cm} R_2\to LR_2L^{\dagger}\ ,
\ee
where $L$ is a $SL(2,C)$ matrix: $L=\left(\begin{array}{cc}
m&n\\
k&l \end{array}\right)$. Note the transformation rule of the functions $V^{-2\eta}(z,\bar{z}; R)$ under $L$:
\be\label{vret}
V^{-2\eta}(z,\bar{z}; LRL^{\dagger})={1\over |nz+l|^{4\eta}}V^{-2\eta}\left({mz+k\over nz+l},c.c; R\right)\, .
\ee
Performing the change of the integration variables   (\ref{rrr}), using
that the $\delta$-function arguments is invariant under  (\ref{rrr}) and  the transformation rule (\ref{vret}), we obtain
\bea
&&\langle V_{\alpha}(z_1,\bar{z}_1)XV_{\alpha}(z_2,\bar{z}_2)\rangle^{\rm light} =\\ \nonumber
&&{1\over |nz_1+l|^{4\eta}}{1\over |nz_2+l|^{4\eta}}
\Big\langle V_{\alpha}\left({mz_1+k\over nz_1+l},c.c.\right)XV_{\alpha}\left({mz_2+k\over nz_2+l},c.c.\right)\Big\rangle^{\rm light}\, ,
\eea
which is the standard consequence of the conformal invariance, when we remember that in the light asymptotic limit
$\Delta_{\eta b}\to\eta$.
This calculation shows that the invariance of the defect parameter $\kappa$ under  (\ref{rrr})
is related to the conformal invariance of the defect two-point function.

Using conformal invariance we can set $z_1$ to $\infty$ and $z_2$ to $0$ to derive:
\bea\label{intjusha}
&&\langle V_{b\eta}(z_1,\bar{z}_1)XV_{b\eta}(z_2,\bar{z}_2)\rangle^{\rm light} ={\tilde{\lambda}^{-2\eta}\over (z_1-z_2)^{2\eta}(\bar{z}_1-\bar{z}_2)^{2\eta}} \\ \nonumber
&\times&\int_{H_3^{+}\times  H_3^{+}} dR_1dR_2
\delta\Big({\rm Tr}\left(R_2R_1^{-1}\right)-2\kappa\Big)
(R_1)_{11}^{-2\eta}(R_2)_{22}^{-2\eta}\, .
\eea

To calculate this integral we  express the Hermitian matrices $R_1$ and $R_2$ as products
\be\label{r1gg}
R_1=gg^{\dagger},\hspace{0.5cm} R_2=\tilde{g}\tilde{g}^{\dagger},\hspace{0.5cm}
g,\tilde{g} \in SL(2,\mathbb{C})\, ,
\ee
implying that
\be\label{vggg1}
V_1=\sqrt{\tilde{\lambda}}\Big(|g_{11}z+g_{21}|^2+
|g_{12}z+g_{22}|^2\Big)\, ,
\ee
\be\label{vggg2}
V_2=\sqrt{\tilde{\lambda}}\Big(|\tilde{g}_{11}z+\tilde{g}_{21}|^2+
|\tilde{g}_{12}z+\tilde{g}_{22}|^2\Big)\, .
\ee
At the next step we will parametrize $\tilde{g}$ as a product of matrices $g$ and $U$:
\be\label{guuu}
\tilde{g}=gU,
\ee
where $U$ is an $SL(2,\mathbb{C})$ matrix
\be
U=\left(\begin{array}{cc}
u_{11}&u_{12}\\
u_{21}&u_{22} \end{array}\right)\, ,\hspace{1cm} u_{11}u_{22}-u_{12}u_{21}=1\, .
\ee

Inserting (\ref{r1gg}) and (\ref{guuu})  in (\ref{11t2})
we obtain
\be\label{kapets}
2\kappa={\rm Tr}\, UU^{\dagger}\, .
\ee


This can be understood noting that the solutions (\ref{vggg1}) and (\ref{vggg2}) correspond to
\be
a_i(z)=g_{1i}z+g_{2i}\, ,\hspace{2cm}\tilde{a}_i(z)=\tilde{g}_{1i}z+\tilde{g}_{2i}\, ,\hspace{2cm} i=1,2\, ,
\ee
\be
b_i(\bar{z})=\bar{g}_{1i}\bar{z}+\bar{g}_{2i}\, ,\hspace{2cm}\tilde{b}_i(\bar{z})=\bar{\tilde{g}}_{1i}\bar{z}+\bar{\tilde{g}}_{2i}\, ,\hspace{2cm} i=1,2\, .
\ee
It is obvious that
\be
\tilde{a}_i=\sum_{j=1}^2a_ju_{ji}\, ,
\ee
\be
\tilde{b}_i=\sum_{j=1}^2b_j\bar{u}_{ji}\, .
\ee
We see that passing from $g$ to $\tilde{g}=gU$ brings to the simultaneous rotations
of $a_i$ and $b_i$, $i=1,2$, by matrices $U$ and $\bar{U}$.
Therefore the defect parameter $\kappa$ is equal to the trace of the product $UU^{\dagger}$.
In these variables the integral (\ref{intjusha}) simplifies and reads
\bea\label{valphak}
&&\langle V_{b\eta}(z_1,\bar{z}_1)XV_{b\eta}(z_2,\bar{z}_2)\rangle^{\rm light} ={\tilde{\lambda}^{-2\eta}\over (z_1-z_2)^{2\eta}   (\bar{z}_1-\bar{z}_2)^{2\eta}} \\ \nonumber
&\times&\int dR_1 dU
\delta(|u_{11}|^2+|u_{12}|^2+|u_{21}|^2+|u_{22}|^2-2\kappa)
(R_1)_{11}^{-2\eta}(R_2)_{22}^{-2\eta}\, ,
\eea
where $dR_1$ and $dU$ are the corresponding integration measures which will be elaborated below.

Using $SU(2)$ freedom in the choice of $g$ we can adopt the parameterization
\be\label{grho}
g=\left(\begin{array}{cc}
\rho_1^{-1}& w_1\\
0&\rho_1 \end{array}\right),\hspace{0.5cm} \rho_1\in \mathbb{R}, \,\,\, w_1\in \mathbb{C}
\ee
and
\be\label{grmr}
R_1=\left(\begin{array}{cc}
\rho_1^{-2}+|w_1|^2&\rho_1 w_1\\
\rho_1 \bar{w}_1&\rho_1^2 \end{array}\right)\, .
\ee

Parameterizing $\tilde{g}$ in the same way
\be
\label{grho2}
\tilde{g}=\left(\begin{array}{cc}
\rho_2^{-1}& w_2\\
0&\rho_2 \end{array}\right),\hspace{0.5cm} \rho_2\in \mathbb{R}, \,\,\, w_2\in \mathbb{C}\, ,
\ee
we find that the elements of the matrix $U=g^{-1}\tilde{g}$ satisfy the relations:
\bea\label{tilka}
&&u_{21}=0\, ,\\ \nonumber
&&u_{22}=u_{11}^{-1}\equiv u\, ,\hspace{1cm} u\in\mathbb{R}\, ,\\ \nonumber
&&\rho_2=\rho_1 u\, ,\\ \nonumber
&&w_2= \rho_1^{-1}u_{12}+w_1u\, .\\
\eea
Eq. (\ref{tilka}) implies
\be
R_2=\left(\begin{array}{cc}
\rho_1^{-2}u^{-2}+|\rho_1^{-1}u_{12}+w_1u|^2&\rho_1u (\rho_1^{-1}u_{12}+w_1u)\\
\rho_1u (\rho_1^{-1}\bar{u}_{12}+\bar{w}_1u)&\rho_1^2 u^2\end{array}\right)\, .
\ee
Using the volume form on the 3D hyperboloid $H_3^+$ computed
in appendix B (\ref{volfor}), one obtains for the integration measure
\be
dR_1dR_2=4\rho_1d\rho_1d^2w_1 udu  d^2u_{12}\, .
\ee
Now the integral  (\ref{valphak}) takes the form
\bea\label{valpha}
&&\langle V_{b\eta}(z_1,\bar{z}_1)XV_{b\eta}(z_2,\bar{z}_2)\rangle^{\rm light} ={4\tilde{\lambda}^{-2\eta}\over (z_1-z_2)^{2\eta}(\bar{z}_1-\bar{z}_2)^{2\eta}} \\ \nonumber
&\times&\int \rho_1d\rho_1d^2w_1 udu  d^2u_{12} \delta\left(u^2+{1\over u^2}+|u_{12}|^2-2\kappa\right)
{1\over (\rho_1^{-2}+|w_1|^2)^{2\eta}}{1\over \rho^{4\eta}_1 u^{4\eta}}\, .
\eea
We see that the delta function in the integrand of (\ref{valpha}) can be different from zero only for $\kappa>1$
in agreement with discussion after formula (\ref{kapka}).
Performing the integral over $u_{12}$
and then over $u$ we obtain
\bea
&&\langle V_{b\eta}(z_1,\bar{z}_1)XV_{b\eta}(z_2,\bar{z}_2)\rangle^{\rm light} =\\ \nonumber
&&2\pi\tilde{\lambda}^{-2\eta}{\Big((\kappa+\sqrt{\kappa^2-1})^{1-2\eta}
-(\kappa-\sqrt{\kappa^2-1})^{1-2\eta}\Big)\over (1-2\eta) (z_1-z_2)^{2\eta}(\bar{z}_1-\bar{z}_2)^{2\eta}} \\ \nonumber
&\times&\int \rho_1d\rho_1d^2w_1
{1\over (\rho_1^{-2}+|w_1|^2)^{2\eta}}{1\over \rho^{4\eta}_1}\, .
\eea
Performing the integral over $w_1$ one gets
\be
\int \rho_1d\rho_1d^2w_1
{1\over (\rho_1^{-2}+|w_1|^2)^{2\eta}}{1\over \rho^{4\eta}_1}
={1\over 2\eta-1}\int{d\rho_1\over \rho_1}={1\over 2\eta-1}\delta(0)\, .
\ee
The integral with respect to $w_1$ converges if $2\eta >1$.
Having computed the integral under this assumption, we can define it away from this region by analytic
continuation.
The integral with respect to $\rho_1$ diverges. This divergence was analyzed in \cite{Seiberg:1990eb} and related to the infinite volume
of the dilation group. It gives rise in fact
to the $\delta(0)$ which appears in the two-point function of coincident fields of the continuous spectrum.
We get a finite result
taking the ratio
\be\label{vbet}
{\langle V_{b\eta}(z_1,\bar{z}_1)XV_{b\eta}(z_2,\bar{z}_2)\rangle^{\rm light}
\over \langle V_{0}(z_1,\bar{z}_1)XV_{0}(z_2,\bar{z}_2)\rangle^{\rm light}}=
{\tilde{\lambda}^{-2\eta}\sinh 2\pi\sigma(1-2\eta)\over
(1-2\eta)^2(z_1-z_2)^{2\eta}(\bar{z}_1-\bar{z}_2)^{2\eta}\sinh 2\pi\sigma}\, .
\ee
Here we set $\kappa=\cosh 2\pi\sigma$.

Using the properties of the $\Gamma$ functions collected in appendix A one can calculate the light asymptotic limit
of the ZZ function (\ref{zzfu}):
\be
{W^{-1}_{\alpha=
\eta b}\over W^{-1}_{\alpha=0}}
\to\tilde{\lambda}^{-\eta}{1\over 1-2\eta}\, ,
\ee
and setting $s={\sigma\over b}$ and $\alpha=\eta b$ we also obtain

\be
{\cosh 2\pi s(2\alpha-Q)\over \cosh 2\pi s Q}\to e^{-4\pi\eta|\sigma|}\, .
\ee
Hence, recalling (\ref{defpoi}) we  get in the light  asymptotic limit for the defect two-point function
derived via the bootstrap program
\be\label{shmet}
{\langle V_{b\eta}(z_1,\bar{z}_1)XV_{b\eta}(z_2,\bar{z}_2)\rangle
\over \langle V_{0}(z_1,\bar{z}_1)XV_{0}(z_2,\bar{z}_2)\rangle}
\to
{\tilde{\lambda}^{-2\eta}\over (2\eta-1)^2}{e^{-4\pi\eta|\sigma|}\over (z_1-z_2)^{2\eta}(\bar{z}_1-\bar{z}_2)^{2\eta}}\, .
\ee

In the limit of  large $\sigma$ we get full agreement between
(\ref{vbet}) and (\ref{shmet}). It may happen that inclusion of one-loop determinants could make this agreement
exact for all values of $\sigma$. The study of this point is left for future work.
It is interesting to note, that in boundary conformal Toda field theory the agreement
between the light asymptotic limit of boundary one-point function with the path integral calculations  was also reached
in \cite{Fateev:2010za} in the limit of the large boundary cosmological constant.

\section{Defects in the heavy asymptotic limit}

\subsection{Heavy asymptotic limit of the correlation functions}

In this section we consider the heavy asymptotic limit of  two-point functions
in the presence of defects (\ref{defpoi}).
Now  we should find asymptotic behaviour  of the inverse ZZ function (\ref{zzfu}) and of the factor
$\cosh(2\pi s (2\alpha-Q))$ in the limit $b\to 0$, setting  $\alpha={\eta\over b}$,
and $s={\sigma\over b}$.
 In  the heavy asymptotic limit we should keep only terms having the form
  $\sim e^{1/b^2}$.

To understand the semiclassical origin of the denominator in (\ref{dalsal}) we find very useful to consider, in the spirit of  \cite{Harlow:2011ny}, analytic continuation
of the Liouville theory with a complex $\eta$ and complex saddle points.


Taking  $\eta$ to satisfy the Seiberg bound (\ref{seibb}) ${\rm Re}\,\eta <{1\over 2}$,
and using properties of $\Gamma$ functions collected in appendix A,
 we obtain


\be\label{wlma}
W^{-1}_{\alpha=
{\eta\over b}}\to C(b,\eta)\lambda^{1-2\eta\over 2b^2}
{1\over \sin \pi \left({2\eta-1\over b^2}\right)}
\exp\left({2\eta-1\over b^2}\Big[\log(1-2\eta)-1\Big]\right)\, .
\ee
where
\bea\label{consneg}
&&C(b,\eta)=-{2^{-3/4}b^3\Gamma(2\eta)\over (2\eta-1)^2}\\ \nonumber
&&=\exp\left(-{3\over 4}\log2+i\pi+\log\Gamma(2\eta)-2\log(1-2\eta)+3\log b\right)
\eea
We see that all the terms in (\ref{consneg}) are  negligible compare to terms growing like
 $\sim e^{1/b^2}$ in the limit $b\to 0$, and therefore  $C(b,\eta)$ can be omitted.
The importance of the term ${1\over \sin \pi \left({2\eta-1\over b^2}\right)}$
is explained in \cite{Harlow:2011ny}.  It was shown  that this term in the
semiclassical interpretation arises as a sum over some ``instanton" like sectors.
As a preparation to this point we will expand this term in two ways as suggested in  \cite{Harlow:2011ny}.
Denoting $y=e^{i\pi(2\eta-1)/b^2}$ one can write
\be
{1\over \sin \pi \left({2\eta-1\over b^2}\right)}={2i\over y-y^{-1}}=2i\sum_{k=0}^{\infty}y^{-(2k+1)}=
-2i\sum_{k=0}^{\infty}y^{2k+1}\, .
\ee
One expansion is valid for $|y|>1$ and one for $|y|<1$. So either way, there is  a set $T$ of integers with
\be\label{qaz}
{1\over \sin \pi \left({2\eta-1\over b^2}\right)}=\pm 2i\sum_{M\in T} e^{2i\pi(M\mp 1/2)(2\eta-1)/b^2}\, ,
\ee
where $T$ consists of nonnegative integers if ${\rm Im}(2\eta-1)/b^2>0$ and of nonpositive ones
if ${\rm Im}(2\eta-1)/b^2<0$.

The set $T$ in (\ref{qaz}) can be understood as sum of saddle points in the minisuperspace approximation keeping only
constant mode of $\phi$. In this approximation the Liouville path integral becomes the integral representation
of the $\Gamma(x)$ function \cite{Harlow:2011ny}:
\be\label{gamint}
 \Gamma(x)=\int_{-\infty}^{\infty}d\phi\exp(-S)\, ,
 \ee
 where the minisuperspace action is
 \be\label{sx}
 S=-x\phi+e^{\phi}\, .
 \ee
 The steepest descent analysis of the $\Gamma(x)$ function
 asymptotic behaviour  for the large negative $x$,
  was carried out  in \cite{boyd}. It is based on the lengthy and careful analysis of the
 integration contours of the integral representation of the $\Gamma(x)$ function (\ref{gamint}), along which it
 converges in quadrants   ${\rm Re}\ x<0$, ${\rm Im}\ x>0$ and  ${\rm Re}\ x<0$, ${\rm Im}\ x<0$. In the physical literature
 it is reviewed in \cite{Harlow:2011ny,Pasquetti:2009jg}. In this way we obtain the factor ${1\over \sin\pi x}$
 in (\ref{gamix}) as a sum over the saddle points of the action (\ref{sx}).

Setting  $\alpha={\eta\over b}$ and $s={\sigma\over b}$ we easily obtain:
\be\label{ealq}
\cosh 2\pi s(2\alpha-Q)\to
e^{{2\pi\over b^2} |\sigma|(1-2\eta)}\, .
\ee

Now we are in a position to write down the limiting form of the defects  correlation functions.

Inserting (\ref{wlma}), (\ref{ealq}) in (\ref{defpoi}) we can write in the heavy asymptotic limit
\bea\label{qwe}
&&\langle V_{\alpha}(z_1,\bar{z}_1)XV_{\alpha}(z_2,\bar{z}_2)\rangle\sim
(z_1-z_2)^{-2\eta(1-\eta)/b^2}(\bar{z}_1-\bar{z}_2)^{-2\eta(1-\eta)/b^2}\\ \nonumber
&\times&\lambda^{1-2\eta\over b^2}
{1\over \sin^2 \pi \left({2\eta-1\over b^2}\right)}
\exp\left({4\eta-2\over b^2}\Big[\log(1-2\eta)-1\Big]\right)e^{{2\pi\over b^2} |\sigma|(1-2\eta)}\, .
\eea
Using also  (\ref{qaz}) we get
\be
\langle V_{\alpha}(z_1,\bar{z}_1)XV_{\alpha}(z_2,\bar{z}_2)\rangle\sim\sum_{M_1,M_2\in T}\exp\left(-S^{\rm def}_{M_1,M_2}\right)\, ,
\ee
where
\bea\label{sdefect}
&&b^2S^{\rm def}_{M_1,M_2}=-2i\pi(M_1+M_2\mp 1)(2\eta-1)+
4\eta(1-\eta)\log|z_1-z_2|\hspace{1cm}\\ \nonumber
&&-(1-2\eta)\log\lambda-
(4\eta-2)\log(1-2\eta)+(4\eta-2)-2\pi|\sigma|(1-2\eta)\, .
\eea
It is instructive to compare the heavy asymptotic limit of the defect two-point function with the corresponding limit
of the  usual two-point function, computed in \cite{Harlow:2011ny}
\bea\label{qwe1}
&&\langle V_{\alpha}(z_1,\bar{z}_1)V_{\alpha}(z_2,\bar{z}_2)
\rangle\sim |z_{1}-z_2|^{-4\eta(1-\eta)/b^2}\\ \nonumber
&\times&
\lambda^{(1-2\eta)/b^2}
{1\over \sin\pi(2\eta-1)/b^2}
\exp\left({4\eta-2\over b^2}[\log(1-2\eta)-1]\right)\, .
\eea
The relation of (\ref{qwe}) to (\ref{qwe1}) naturally gives the heavy asymptotic limit
of the eigenvalues ${\cal D}_s(\alpha)$ of the defect operator:
\be\label{dalpha}
{\cal D}_s(\alpha)={\langle V_{\alpha}(z_1,\bar{z}_1)XV_{\alpha}(z_2,\bar{z}_2)\rangle
\over \langle V_{\alpha}(z_1,\bar{z}_1)V_{\alpha}(z_2,\bar{z}_2)
\rangle}\to{e^{{2\pi\over b^2} |\sigma|(1-2\eta)}\over \sin \pi \left({2\eta-1\over b^2}\right)}\, .
\ee
Of course  (\ref{dalpha}) can be also easily derived directly from (\ref{dalsal}) in the heavy asymptotic limit.

\subsection{Evaluation of the action for classical solutions}
According to the general prescription of the semiclassical heavy asymptotic limit, we
should find solutions of the Liouville equation, satisfying the defect equations of motion
and possessing the logarithmic singularities (\ref{viscon}) at points $z_1$ and $z_2$.
 The form of the solution of the defect equations of motion  (\ref{phi1}) and (\ref{phi2}) implies
that we should find functions $A(z)$, $C(z)$ and $B(\bar{z})$ in such a way that $\phi_1$ has a logarithmic
singularity at the point $z_1$ and $\phi_2$ has a logarithmic
singularity at the point $z_2$. Since the energy-momentum tensor is continuous across a defect this implies
that we should find solutions possessing two singular points.
Two-point solutions are well known ( see for example \cite{Harlow:2011ny}) and we can build from them the Ansatz
satisfying the defect equations of motion.

To build the solution with the required singularities one should take a function $A(z)$
which is smooth and holomorphic away from $z_1$ and $z_2$.
Let us take $A(z)$ as
\be\label{aiki}
A(z)=e^{2\nu_1}(z-z_1)^{2\eta-1}(z-z_2)^{1-2\eta}\, .
\ee
One has also
\be\label{aeta}
a_1={1\over \sqrt{\partial A}}={e^{-\nu_1}\over \sqrt{(z_1-z_2)(2\eta-1)}}(z-z_1)^{1-\eta}(z-z_2)^{\eta}\, ,
\ee
\be\label{abeta}
a_2={A\over \sqrt{\partial A}}={e^{\nu_1}\over  \sqrt{(z_1-z_2)(2\eta-1)}}(z-z_1)^{\eta}(z-z_2)^{1-\eta}\, .
\ee
Inserting (\ref{aeta}) or (\ref{abeta}) in (\ref{tbaa}) we obtain the energy-momentum tensor
\be
b^2T={\eta(1-\eta)\over (z-z_1)^2}+{\eta(1-\eta)\over (z-z_2)^2}-
{2\eta(1-\eta)\over z_1-z_2}\left({1\over z-z_1}-{1\over z-z_2}\right)\, ,
\ee
which indeed possesses two singular points (\ref{tc12}), with accessory parameters
\be\label{ceta}
c_2=-c_1={2\eta(1-\eta)\over z_1-z_2}\, .
\ee
The anti-holomorphic part is:

\be\label{zaiki}
B(\bar{z})=-(\bar{z}-\bar{z}_1)^{1-2\eta}(\bar{z}-\bar{z}_2)^{2\eta-1}\, ,
\ee
\be\label{abetab}
b_1={B\over \sqrt{\bar{\partial} B}}={1\over  \sqrt{(\bar{z}_1-\bar{z}_2)(2\eta-1)}}(\bar{z}-\bar{z}_1)^{1-\eta}(\bar{z}-\bar{z}_2)^{\eta}\, ,
\ee
\be\label{abetab1}
b_2=-{1\over \sqrt{\bar{\partial} B}}={1\over  \sqrt{(\bar{z}_1-\bar{z}_2)(2\eta-1)}}(\bar{z}-\bar{z}_1)^{\eta}(\bar{z}-\bar{z}_2)^{1-\eta}\, .
\ee
Let us take the holomorphic part for $\phi_2$ as
\be\label{aiki0}
C(z)=e^{2\nu_2}(z-z_1)^{2\eta-1}(z-z_2)^{1-2\eta}=e^{2(\nu_2-\nu_1)}A(z)\, ,
\ee
and the antiholomorphic part again given by (\ref{zaiki}).
Using (\ref{aldet}) one gets
\be\label{coshik}
\kappa=\cosh(\nu_2-\nu_1)\, .
\ee
Inserting (\ref{aiki}), (\ref{aiki0}) and  (\ref{zaiki}) in (\ref{phi1}) and (\ref{phi2})  we obtain:
\bea\label{solod1}
e^{-\varphi_1}={\lambda
\over (2\eta-1)^2|z_1-z_2|^2}
\Big(
e^{\nu_1}|z-z_1|^{2\eta}|z-z_2|^{2-2\eta}\\ \nonumber
-e^{-\nu_1}
|z-z_1|^{2-2\eta}|z-z_2|^{2\eta}\Big)^2\, ,
\eea
\bea\label{solod2}
e^{-\varphi_2}={\lambda
\over (2\eta-1)^2|z_1-z_2|^2}
\Big(
e^{\nu_2}|z-z_1|^{2\eta}|z-z_2|^{2-2\eta}\\ \nonumber
-e^{-\nu_2}
|z-z_1|^{2-2\eta}|z-z_2|^{2\eta}\Big)^2\, .
\eea
It is easy to see that $\varphi_1$ and $\varphi_2$ given by (\ref{solod1}) and (\ref{solod2}) have the required singularity (\ref{viscon}) around $z_1$
and  $z_2$ respectively.
In fact each of the functions $\varphi_1$ or $\varphi_2$ given by  (\ref{solod1}) and (\ref{solod2}) coincides
 with the solution describing a saddle point
for a two-point function considered in  \cite{Harlow:2011ny}. But in \cite{Harlow:2011ny}
this solution was considered on a full plane with the same parameter $\nu$ everywhere, whereas
 here each of them is considered on a corresponding half-plane,
namely in (\ref{solod1}) $z$ belongs to the upper half-plane $\Sigma_1$, and in (\ref{solod2}) $z$
belongs to the  lower half-plane $\Sigma_2$,
and we should also remember that,
 $z_1\in \Sigma_1$
and   $z_2\in\Sigma_2$.
 The  defect is created
by the choice of different parameters $\nu_1$ and $\nu_2$, $\nu_1\neq \nu_2$.

From (\ref{solod1}) and (\ref{solod2}) we obtain
\bea\label{solod11}
&&\varphi_1=4i\pi N_1-\log\lambda+2\log(1-2\eta)\\ \nonumber
&&-2\log\left(
{e^{\nu_1}|z-z_1|^{2\eta}|z-z_2|^{2-2\eta}\over |z_1-z_2|}-
{e^{-\nu_1}
|z-z_1|^{2-2\eta}|z-z_2|^{2\eta}\over |z_1-z_2|}\right)\, ,
\eea
\bea\label{solod22}
&&\varphi_2= 4i\pi N_2-\log\lambda+2\log(1-2\eta)\\ \nonumber
&&-2\log\left(-
{e^{\nu_2}|z-z_1|^{2\eta}|z-z_2|^{2-2\eta}\over |z_1-z_2|}+
{e^{-\nu_2}
|z-z_1|^{2-2\eta}|z-z_2|^{2\eta}\over |z_1-z_2|}\right)\, .
\eea
Here $N_1$ and $N_2$ are integer. The possibility to add the term $4i\pi N_j$, $j=1,2$, results from the invariance
of the bulk (\ref{leom}) and defect (\ref{def3})-(\ref{def111}) Liouville equations of motion under the transformation
$\phi_j\to \phi_j+2\pi i N_j/b$, or multiplying by $2b$, under
 $\varphi_j\to \varphi_j+4\pi i N_j$, $j=1,2$. Note that the bulk Liouville equation (\ref{leom})
is invariant under the  symmetry $\varphi_j\to \varphi_j+2\pi i N_j$, and it is broken to  $\varphi_j\to \varphi_j+4\pi i N_j$
by the exponential terms of the defect action (\ref{topdef}).

To evaluate the action on the solutions (\ref{solod1}),  (\ref{solod2}), we will use the strategy used in
\cite{Zamolodchikov:1995aa}. Namely we will write the system of differential equations
which this action should satisfy.
The first equation is (\ref{zamequ}), which given that $\eta_1=\eta_2=\eta$, reads
\be\label{sx1x2}
b^2{\partial S^{\rm def}_{\rm cl}\over \partial\eta}=-X_1-X_2\, .
\ee
where $X_i$ is defined in (\ref{viscon}).
The leading terms of $\varphi_1$ around $z_1$ are
\be
\varphi_1\to -4\eta\log|z-z_1|+X_1\, ,
\ee
where
\be\label{xx11}
X_1=4\pi i N_1-\log\lambda+2\log(1-2\eta)-(2-4\eta)\log|z_1-z_2|-2\nu_1\, .
\ee

Similarly the leading terms of $\varphi_2$ around $z_2$  are
\be
\varphi_2\to -4\eta\log|z-z_2|+X_2\, ,
\ee
where
\be\label{xx22}
X_2=4\pi i N_2-\log\lambda+2\log(1-2\eta)-(2-4\eta)\log|z_1-z_2|+2\nu_2\, .
\ee
Inserting (\ref{xx11}) and (\ref{xx22}) in (\ref{sx1x2}) one obtains
\be\label{sx12}
b^2{\partial S^{\rm def}_{\rm cl}\over \partial\eta}=
-2\pi i \left(2N_1+2N_2\right)+2\log\lambda-4\log(1-2\eta)+(4-8\eta)\log|z_1-z_2|+2(\nu_1-\nu_2)\, .
\ee
We would like to emphasize yet another difference from the calculation of the heavy asymptotic limit of the two-point function in \cite{Harlow:2011ny}.
In the case of the usual two-point function the integers $N_1$ and $N_2$ are equal since we have one continuous function $\phi$.
Here they can be different since we have two different functions $\varphi_1$ and $\varphi_2$.

The action with defect (\ref{topdefbb}) implies also
\be\label{skap}
b^2{\partial S^{\rm def}_{\rm cl}\over \partial \kappa}={1\over i\pi}\int_{\partial \Sigma_1} e^{\Lambda b}d\tau\, .
\ee
Inserting (\ref{aiki}) and (\ref{aiki0}) in eq. (\ref{lambdik}) one obtains
\be\label{lambdon}
e^{\Lambda b}={1\over 2\sinh(\nu_1-\nu_2)}{(2\eta-1)(z_1-z_2)\over (z-z_1)(z-z_2)}\, .
\ee
Using that
\be
{1\over i}\int_{\partial \Sigma_1}{dz\over (z-z_1)(z-z_2)}={2\pi\over (z_1-z_2)}\, ,
\ee
we obtain
\be\label{skappa}
b^2{\partial S^{\rm def}_{\rm cl}\over \partial \kappa}={2\eta-1\over \sinh(\nu_1-\nu_2)}\, .
\ee

Integrating equations (\ref{sx12}) and (\ref{skappa}) we obtain:
\bea\label{sdefecter}
&&b^2S^{\rm def}_{N_1,N_2}=-2i\pi(2N_1+2N_2)\eta+
4\eta(1-\eta)\log|z_1-z_2|\\ \nonumber
&&+2\eta\log\lambda-
(4\eta-2)\log(1-2\eta)+4\eta-(\nu_1-\nu_2)(1-2\eta)+C\, ,
\eea
where $C$ is a constant.
To derive the penultimate term we should remember the relation (\ref{coshik}).
To fix the constant term we can directly compute the action (\ref{topdefbb})  for the Ansatz  (\ref{solod11})-(\ref{solod22}) with
$\eta=0$:

\be\label{valk1}
\varphi_1=4i\pi N_1-\log\lambda-\log\left({e^{\nu_1}\over |z_1-z_2|}|z-z_2|^2-
{e^{-\nu_1}\over |z_1-z_2|}|z-z_1|^2\right)^2\, ,
\ee

\be\label{valk2}
\varphi_2=4i\pi N_2-\log\lambda-\log\left({e^{\nu_2}\over |z_1-z_2|}|z-z_2|^2-
{e^{-\nu_2}\over |z_1-z_2|}|z-z_1|^2\right)^2\, .
\ee



Evaluation of the action (\ref{topdefbb}) on the Ansatz (\ref{valk1}), (\ref{valk2})
is lengthy and explained in appendix C. The result is

\be\label{bs}
b^2S_0=2i\pi(N_1+N_2)-\log \lambda-2-(\nu_1-\nu_2)\, .
\ee

Comparing (\ref{bs}) with (\ref{sdefecter}) fixes the constant $C$:
\be
C=2i\pi(N_1+N_2)-\log \lambda-2\, .
\ee
Inserting this value of  $C$ in  (\ref{sdefecter}) we indeed obtain (\ref{sdefect})
if we set
\be
N_1=M_1\, ,
\ee
\be
N_2=M_2\mp 1\, ,
\ee
and
\be
2\pi\sigma=\nu_1-\nu_2\, .
\ee

Some comments are in order at this point:
\begin{enumerate}
\item
The action (\ref{sdefecter}) satisfies the Polyakov relation (\ref{polyakus}) with the accessory parameters
defined in (\ref{ceta}):
\be\label{spolik}
b^2{\partial S^{\rm def}_{\rm cl}\over \partial z_i}=(-)^{i+1}{2\eta(1-\eta)\over z_1-z_2}\, ,\hspace{1cm} i=1,2\, .
\ee
\item
In eq.\@ (\ref{qaz}) $M$ takes nonnegative integer values if ${\rm Im}(2\eta-1)/b^2>0$,
and nonpositive integer values if ${\rm Im}(2\eta-1)/b^2<0$. Therefore $N_1$ also runs over
nonnegative or nonpositive integer values depending on the sign of ${\rm Im}(2\eta-1)/b^2$,
and $N_2$ takes values $\{1,2,\ldots\}$, when  ${\rm Im}(2\eta-1)/b^2>0$ and
$N_2$ takes values $\{-1,-2,\ldots\}$, when  ${\rm Im}(2\eta-1)/b^2<0$.
The fact that for the different values of the parameter $\eta$ we should take contribution of  different sets of the saddle points
is known as the Stokes phenomena \cite{Witten:2010cx,Harlow:2011ny,Marino:2012zq,Pasquetti:2009jg}, and was studied in detail for two- and three-point correlation functions
 of the Liouville field theory in \cite{Harlow:2011ny}.
 Recall that it is caused by the fact that the sum  (\ref{qaz}) converges for
 the different values of $M$ depending on the sign of ${\rm Im}(2\eta-1)/b^2$.
 The values of parameters at which the jump of the set of the contributing saddle point occurs define a (anti-) Stokes line.
 We have a Stokes line if at some values of parameters
the imaginary parts of the  actions for two saddle points, say $a$ and $b$, coincide:
${\rm Im}\ S_{a}={\rm Im}\ S_{b}$.  We have an anti-Stokes line if at some values of parameters the
real parts of the actions for two saddle points, say $a$ and $b$, coincide:
${\rm Re}\ S_{a}={\rm Re}\ S_{b}$. Crossing these lines, a jump in the set
of the contributing saddle point may occur. For the Stokes lines it is caused by the fact that
there is a steepest descent contour connecting two saddle points. For the anti-Stokes line it
is implied by the coincidence of the magnitudes of the amplitudes $e^{S_a}$ and $e^{S_b}$ for the different saddle points.
 From (\ref{sdefect}) or (\ref{sdefecter}) we see that ${\rm Re}\;S^{\rm def}_{N_1,N_2}$ are the same
 for all $N_1$ and $N_2$ if ${\rm Im}(2\eta-1)=0$.
 The line ${\rm Im}(2\eta-1)=0$ is the anti-Stokes line at which indeed we observe a jump in the set
 of the contributing saddle points.

\item
The discussion above of the differences between the calculation of two-point
function with and without defect
suggests nice interpretation of the defect operator.
We have seen that there exist two sources of discontinuity giving rise to the corresponding
terms in the defect operators.
The heavy asymptotic limit of $D(\alpha)$ (\ref{dalpha}) has an exponential in the numerator and sine function in the denominator.
The exponential term in the numerator as we have seen originates from the discontinuity created by the choice of
the different parameters $\nu_1$ and $\nu_2$.
The correspondence between the $N_i$ and $M_i$  parameters makes clear that the different logarithmic branch
solutions, given by $N_1$ and $N_2$, are responsible for the quadratic  $\sin^2 \pi \left({2\eta-1\over b^2}\right)$ term in the (\ref{qwe}).
On the  other hand, as we have mentioned before, in the heavy asymptotic limit the calculation of the usual two-point function
one has $N_1=N_2$, and it reflects the presence of the term
$\sin \pi \left({2\eta-1\over b^2}\right)$  in the denominator of (\ref{qwe1}) in the first degree.
Therefore the denominator  $\sin \pi \left({2\eta-1\over b^2}\right)$ in
$D(\alpha)$
reflects  the possibility
of the choice of different logarithmic branches with $N_1\neq N_2$ in the solution of the defect equations of motion.
The final quantum expression (\ref{dalsal}) results from the quantum corrections restoring $b\leftrightarrow b^{-1}$
duality of the Liouville theory.
\end{enumerate}
Let us analyze in the heavy asymptotic limit also the relation  (\ref{relaa}) between parameter $s$ and $A(b)$

\be\label{relaaq}
2\cosh 2\pi bs=A(b)\left({W(-b/2)\over W(0)}\right)^2\, .
\ee
It is easy to compute that
\be
\rm{lim}_{b\to 0}{W(-b/2)\over W(0)}=-{2\over\sqrt{\lambda}}\, .
\ee

Setting that $s={\sigma\over b}$, we get
\be
 \cosh 2\pi \sigma={2A(0)\over\lambda}\, .
\ee
This implies that the parameter $\kappa$ is proportional to $A(0)$:
\be
\kappa={2A(0)\over\lambda}\, .
\ee
Note that  in the light asymptotic limit as well as in the heavy asymptotic limit we get the same relation between $\sigma$ and $\kappa$
\be
\kappa=\cosh 2\pi\sigma\, .
\ee

\section{Discussion}

The methods developed in this paper can be applied  to other theories with defects, like
$N=1$ superconformal Liouville theory, conformal and superconformal Toda theories.

The Lagrangian of the $N=1$ Liouville theory with defects is constructed in \cite{Aguirre:2013zfa}
using the technique of the type II integrable defects.
The defect two-point functions in  superconformal Liouville theory can  as well be constructed via the bootstrap program \cite{hasgor}.
It is interesting to use the methods of this paper to
construct solutions for superconformal Liouville field theory  of the defect equations of motion and
study the light and heavy asymptotic limits.

The defect operators in conformal Toda field theory are constructed in \cite{Drukker:2010jp,Sarkissian:2011tr}.
It is possible using methods of this paper together with the technique of type II defects to construct the
Lagrangian of conformal Toda field theory with topological defects and compare with semiclassical limits
of defect two-point functions. This program can as well  be generalized to superconformal Toda field theory.

Let us mention also other interesting problems where methods developed in this paper can be applied.

One of the most important problems regarding non-rational conformal field theories is to find  for them
a relation to a three-dimensional topological field theory description similar to that of the rational ones.
This is still a rather difficult and poorly studied problem. The first step was done in \cite{Verlinde:1989ua}, where the classical phase space of the Chern-Simons gauge
theory with $SL(2,\mathbb{R})$ gauge group has been studied and shown to coincide with the  Teichm\"{u}ller
space of Riemann surfaces. It is established by now \cite{Teschner:2003at}, that the Hilbert space of states obtained by quantizing
the Teichm\"{u}ller space is isomorphic to the space of conformal blocks of Liouville theory.
The methods and solutions derived in this paper can be useful to elaborate on the relation between Chern-Simons gauge theory,
Teichm\"{u}ller
space of Riemann surfaces, and Liouville field theory
 including also  defects.

Defects appear in many areas in  String theory as well as in condensed matter. In particular
they play an important role in the entropy entanglement problems \cite{Calabrese:2004eu}. The methods of semiclassical calculations
of the defect two-point functions developed here can be used also in that areas.
 As we mentioned in the introduction heavy and light asymptotic limits appear in many instances of AGT and AdS/CFT
correspondences. The insights gained in the study of these limits in the presence of defects can be useful
to incorporate defects in these problems.

\section*{Acknowledgments}
The work of H.P. was partially supported by the Armenian SCS  grant 13-1C132
and by the Armenian-Russian SCS grant-2013.
The work of G.S. was partially supported by the Armenian SCS  grant 13-1C278 and
ICTP Network NET68.
We would like to thank Rubik Poghossian and Shmuel Elitzur for many useful discussions.
We thank also the JHEP referee for his valuable comments and suggestions which contributed
to the substantial improvement of the text of the paper.
\\
\vspace*{3pt}

\appendix
\section{Properties of $\Gamma$ functions}

The limiting behavior of the terms with $\Gamma$ functions can be calculated
using the approximation
\be
\Gamma(x)\sim e^{x\log x-x+O(\log x)}\, .
\ee
for $x$ with large positive real part.

For $x$ with negative real part using the formula
\be\label{gamix}
\Gamma(x)\Gamma(-x)=-{\pi\over x\sin\pi x}\, ,
\ee
 one can
bring problem to the previous case.

We also need the well-known behavior of the $\Gamma(x)$ function for $x$ around zero:
\be
\Gamma(x)\sim {1\over x}\, .
\ee
\section{Volume form on the 3D hyperboloid $H_3^{+}$}
The 3D hyperboloid $H_3^{+}$ is a pseudo-sphere
\be
X_0^2-X_1^2-X_2^2-X_3^2=1
\ee
in the ambient Minkowski space with the metric:
\be\label{ambik}
ds^2=-dX_0^2+dX_1^2+dX_2^2+dX_3^2\, .
\ee
In the parametrization (\ref{grmr}),
one has
\bea\label{x0x1x2}
&&X_0-X_1={1\over \rho^2}+|w|^2\, ,\\ \nonumber
&&X_0+X_1=\rho^2\, ,\\ \nonumber
&&X_2+iX_3=\rho w\, ,\\ \nonumber
&&X_2-iX_3=\rho \bar{w}\, .
\eea
Substituting  (\ref{x0x1x2}) in (\ref{ambik}) one obtains
\be
ds^2=4{d\rho^2\over \rho^2}+\rho^4d\left(w\over \rho\right)d\left(\bar{w}\over \rho\right)\, .
\ee
The corresponding volume form is
\be\label{volfor}
\sqrt{{\rm det} G}d\rho d^2w=2\rho d\rho d^2w\, .
\ee

\section{Action evaluation}

The solutions (\ref{valk1}) and (\ref{valk2}) have the form:
\bea
&&\varphi_1=4i\pi N_1-\log\lambda-2\log Z_1\, ,\\ \nonumber
&&\varphi_2=4i\pi N_2-\log\lambda-2\log Z_2\, ,
\eea
where
\bea
&&Z_1=s_1z\bar{z}+t_1z+u_1\bar{z}+v_1\, ,\\ \nonumber
&&Z_2=s_2z\bar{z}+t_2z+u_2\bar{z}+v_2\, ,
\eea
with
\begin{eqnarray}\label{stuv}
&&s_j=\pm\frac{2\sinh\nu_j}{|z_1-z_2|},\quad
u_j=\pm\frac{e^{-\nu_j} z_1-e^{\nu_j} z_2}{|z_1-z_2|},\\ \nonumber
&&t_j=\pm\frac{e^{-\nu_j} \bar{z}_1-e^{\nu_j} \bar{z}_2}{|z_1-z_2|},\quad
v_j=\pm\frac{e^{\nu_j}|z_2|^2-e^{-\nu_j} |z_1|^2}{|z_1-z_2|}\, , \quad j=1,2\, .
\end{eqnarray}
where we take upper signs for $\nu_j$ positive and lower signs for $\nu_j$ negative.
This choice of signs makes $s_j\geq 0$.
Note that
\be
s_jv_j-u_jt_j=-1\, \quad j=1,2\, .
\ee
It is useful to introduce also real and imaginary parts of $u_i$ and $t_i$:
\be
t_j=m_j+in_j\, ,\quad u_j=m_j-in_j\, ,\quad  j=1,2\, .
\ee
The function $\tilde{\Lambda}$ can be found setting $\eta=0$ in
(\ref{lambdon})
\be
e^{-\tilde{\Lambda}/2}={2\sinh(\nu_2-\nu_1)\over z_1-z_2}(z-z_1)(z-z_2)\, .
\ee
Before starting the calculations one should examine the zeros of $Z_1$ and $Z_2$.
It is easy to see, that $Z_j$ as a quadratic form, vanishes on a circle $C_j$ with the center $\left(-{m_j\over s_j},{n_j\over s_j}\right)$ and the radius ${1\over s_j}$, $j=1,2$. Since we have the topological defect, as long
as the discs confined by $C_1$ and $C_2$ do not overlap, we can avoid singularities moving the
defect to the safe region between $C_1$ and $C_2$.
Remember that the defect is located along the horizontal axis, and $\varphi_1$ and $\varphi_2$ are considered on the upper and lower half-planes respectively.
 Therefore $Z_1$ has no zeros if $C_1$ is located in the lower half-plane and $Z_2$ has no zeros if $C_2$ is located in the upper half-plane.
This happens, when
\be\label{nn1}
n_1<-1
\ee
and
\be\label{nn2}
n_2>1\, .
\ee
These constraints enable us to avoid the  singularities.

Check when these constraints are satisfied. Writing $z_1=x_1+iy_1$, and $z_2=x_2+iy_2$, we get
from eq. (\ref{stuv}):
\be
n_j=\pm\left({e^{\nu_j}y_2-e^{-\nu_j}y_1\over |z_2-z_1|}\right)\, .
\ee
Recalling that $y_1>0$, and $y_2<0$, and that we should take upper signs for positive $\nu_j$ and lower sign for negative $\nu_j$,
we see that we obtain negative $n_1$ and positive $n_2$ taking
\be
\nu_1>0\, , \quad {\rm and}\quad n_1={e^{\nu_1}y_2-e^{-\nu_1}y_1\over |z_2-z_1|}\, ,
\ee
\be
\nu_2<0\, , \quad {\rm and}\quad n_2={e^{-\nu_2}y_1-e^{\nu_2}y_2\over |z_2-z_1|}\, .
\ee
Taking $|\nu_j|$ big enough we can always satisfy the condition $|n_j|>1$.
This means also that we take in (\ref{stuv}) the upper sign for $j=1$ and the lower sign $j=2$.


Let us now insert the solution (\ref{valk1}) and (\ref{valk2}) in the action (\ref{topdefbb}).
We will evaluate each term in the $R\to \infty$ limit.
Start by computing the bulk part. The bulk Lagrangians can be written as  a total derivative:
\be\label{pvar}
{1\over 8\pi i}\left(\partial\varphi_j\bar{\partial} \varphi_j+4\lambda e^{\varphi_j}\right)=\partial_{{\bar z}}K^j_z-
\partial_zK^j_{\bar{z}}\, ,\quad j=1,2\, ,
\ee
where
\bea\label{kjz}
K^j_z={1\over 4\pi i}\left(-\frac{2}{ (s_j z+u_j) (s_j |z|^2+t_j z
+u_j \bar{z}+v_j)}
+\frac{s_j \log  (s_j |z|^2 +t_jz
+u_j \bar{z}+v_j)}{ (s_j z+u_j)}\right)\, ,\hspace{1cm}
\eea
and $K^j_{\bar{z}}=\bar{K^j_z}$.
We see that under the conditions (\ref{nn1}) and (\ref{nn2}) the denominators in (\ref{kjz}) have no singular points.

The integral over the r.h.s. of (\ref{pvar}) reduces to the contour integral:
\be
\int_{\Sigma^R_j}\left(\partial_{{\bar z}}K^j_z-
\partial_zK^j_{\bar{z}}\right)d^2 z
=\int_{\mathbb{R}} d\tau (K^j_z+K^j_{\bar{z}})+\int_{{s_R}_j}(iK^j_zz-iK^j_{\bar{z}}\bar{z})d\theta\, .
\ee
The integral over the semi-circle of the big radius $R$ is evaluated to yield
\be
\int_{{s_R}_j}(iK^j_zz-iK^j_{\bar{z}}\bar{z})d\theta={1\over 2}\log s_j R^2\, .
\ee
On the other hand the regularizing terms in the action produce
\be
{1\over 2\pi}\int_{{s_R}_1}\varphi_1 d\theta+\log R=-\log \left(s_1 R^2\right)+\log R+2i\pi N_1-{1\over 2}\log\lambda\, ,
\ee
\be
{1\over 2\pi}\int_{{s_R}_2}\varphi_2 d\theta+\log R=-\log \left(s_2 R^2\right)+\log R+2i\pi N_2-{1\over 2}\log\lambda\, .
\ee
The integral over the real axis of the first term  in $K^j$ gives
\be
-{1\over 2\pi }\int d\tau\frac{2n_j}{ (s_j \tau+u_j)(s_j \tau+t_j) (s_j \tau^2+2m_j \tau
+v_j)}=-{n_j\over \sqrt{n_j^2-1}}+{\rm sgn}(n_j)
\ee
and of the second produces
\be
{1\over 2\pi }\int d\tau\frac{n_js_j \log  (s_j \tau^2 +2m_j\tau
+v_j)}{ (s_j \tau+u_j)(s_j \tau+t_j) }=-\log\left[{\rm sgn}(n_j)\left(n_j- \sqrt{n_j^2-1}\right)\right]-{1\over 2}{\rm sgn}(n_j)\log s_j\, .
\ee
Here we introduced the sign function ${\rm sgn}(x)\equiv{x\over |x|}$.

Remembering that for $\varphi_1$ and $\varphi_2$ the integrals over the real axis run
in the opposite directions, we obtain  finally:
\bea\label{shedu}
&&{1\over 8\pi i}\int_{\Sigma_1^R}\left(\partial\varphi_1\bar{\partial} \varphi_1+4\lambda e^{\varphi_1}\right)d^2 z
+{1\over 8\pi i}\int_{\Sigma_2^R}\left(\partial\varphi_2\bar{\partial} \varphi_2+4\lambda e^{\varphi_2}\right)d^2 z\quad\\ \nonumber
&&+{1\over 2\pi}\int_{{s_R}_1}\varphi_1 d\theta+\log R
+{1\over 2\pi}\int_{{s_R}_2}\varphi_2 d\theta+\log R=\\ \nonumber
&&-\frac{n_1}{\sqrt{n_1^2-1}}+\frac{n_2}{\sqrt{n_2^2-1}}
+\log\left[\frac{n_2-\sqrt{n_2^2-1}}{-n_1+\sqrt{n_1^2-1}}\right]\\ \nonumber
&&+2i\pi N_1+2i\pi N_2-\log\lambda-2\, .
\eea

Now we turn to the calculation of the integrals living on the defect.
The sum of the last two terms in  (\ref{topdefbb}), according to the equations of motion
(\ref{def3}) and (\ref{kapik}) is
\begin{eqnarray}\label{lamisu}
&&-\frac{1}{4\pi i}\int_{-\infty}^{\infty}\mathrm{d}\tau (\partial-\bar{\partial})(\varphi_1-\varphi_2)=\\ \nonumber
&&\frac{1}{\pi }\int_{-\infty}^{\infty}\mathrm{d}\tau\left({n_1\over s_1\tau^2+2m_1\tau+v_1}-
{n_2\over s_2\tau^2+2m_2\tau+v_2}\right)=\\ \nonumber
&&\frac{n_1}{\sqrt{n_1^2-1}}-\frac{n_2}{\sqrt{n_2^2-1}}\, .
\end{eqnarray}

We see that (\ref{lamisu}) cancels the first two terms in the third line of (\ref{shedu}).

Now let us compute the second term on the defect:
\begin{eqnarray}
&&\frac{1}{8\pi i}\int_{-\infty}^{\infty}\mathrm{d}\tau \tilde{\Lambda} (\partial+\bar{\partial})(\varphi_1-\varphi_2)=\hspace{3cm}\nonumber \\
&&{1\over \pi i}\int_{-\infty}^{\infty}\mathrm{d}\tau\log\left[(\tau-z_1)(\tau-z_2)\right]\left({s_1\tau+m_1\over s_1\tau^2+2m_1\tau+v_1}
-{s_2\tau+m_2\over s_2\tau^2+2m_2\tau+v_2}\right)\hspace{1cm}\\ \nonumber
&=&\log\left[{\frac{-m_1+i\sqrt{n_1^2-1}-z_2 s_1}{m_1+i\sqrt{n_1^2-1}+z_1 s_1}}\right]-
\log\left[{\frac{-m_2+i\sqrt{n_2^2-1}-z_2 s_2}{m_2+i\sqrt{n_2^2-1}+z_1 s_2}}\right]\, .
\end{eqnarray}

To simplify this expression one can show, introducing an angle $e^{i\xi}={z_2-z_1\over |z_2-z_1|}$, that
\bea
&&-m_j+i\sqrt{n_j^2-1}-z_2 s_j=i\left((-)^jie^{-\nu_j}e^{i\xi}-n_j+\sqrt{n_j^2-1}\right)\, ,\\ \nonumber
&&m_j+i\sqrt{n_j^2-1}+z_1 s_j=i\left((-)^{j+1}ie^{\nu_j}e^{i\xi}+n_j+\sqrt{n_j^2-1}\right)\, ,\quad j=1,2\, .
\eea
We can also prove
\bea\label{njnj}
\left((-)^jie^{-\nu_j}e^{i\xi}-n_j+\sqrt{n_j^2-1}\right)\left((-)^{j+1}ie^{\nu_j}e^{-i\xi}-n_j+\sqrt{n_j^2-1}\right)\, ,\\ \nonumber
=(-)^j\left(-n_j+\sqrt{n_j^2-1}\right)\left(ie^{-\nu_j}e^{i\xi}-ie^{\nu_j}e^{-i\xi}-(-)^j2n_j\right)\, ,
\eea
and writing $z_1=x_1+iy_1$ and $z_2=x_2+iy_2$,
one obtains that
\be\label{ejnu}
\left(ie^{-\nu_j}e^{i\xi}-ie^{\nu_j}e^{-i\xi}-2(-)^jn_j\right)=-{2i\sinh \nu_j\over |z_2-z_1|}\left(x_2-x_1+i(y_2+y_1)\right)\, .
\ee
And finally we need
\bea
&&\left((-)^{j+1}ie^{\nu_j}e^{-i\xi}-n_j+\sqrt{n_j^2-1}\right)\left((-)^{j+1}ie^{\nu_j}e^{i\xi}+n_j+\sqrt{n_j^2-1}\right)\hspace{1cm}\\ \nonumber
&&=-2e^{\nu_j}\Big(\cosh\nu_j+(-)^ji\cos\xi\sqrt{n_j^2-1}+(-)^jn_j\sin\xi\Big)\, .
\eea
Using all these identities, and noting that the terms in the r.h.s. of
(\ref{ejnu}), independent on $j$,  get canceled, we obtain
\bea\label{purfaso}
&&\frac{1}{8\pi i}\int_{-\infty}^{\infty}\mathrm{d}\tau \tilde{\Lambda} (\partial+\bar{\partial})(\varphi_1-\varphi_2)=\\ \nonumber
&&\log\left[{\Big(\cosh\nu_2+i\cos\xi\sqrt{n_2^2-1}+n_2\sin\xi\Big)\sinh\nu_1\left(-n_1+\sqrt{n_1^2-1}\right) \over \Big(\cosh\nu_1-i\cos\xi\sqrt{n_1^2-1}-n_1\sin\xi\Big)\sinh\nu_2\left(n_2-\sqrt{n_2^2-1}\right)}\right]+\nu_2-\nu_1\, .
\eea
The third multipliers in the numerator and in the denominator of the argument of the logarithm
in (\ref{purfaso}) together
cancel the third term in the third line of (\ref{shedu}).
It is easy to see that the remaining  logarithmic term after this cancellation is a pure argument since the remaining numerator and denominator have the same modulus:
\be
{1\over \sinh^2\nu_j}\Big[(\cosh\nu_j+(-)^jn_j\sin\xi)^2+(n_j^2-1)\cos^2\xi\Big]={(x_1-x_2)^2+(y_1+y_2)^2\over |z_1-z_2|^2}\, .
\ee
And finally the first integral on the defect is
\begin{eqnarray}\label{purfase}
&&-\frac{1}{16\pi i}\int_{-\infty}^{\infty}\mathrm{d}\tau(\varphi_2 \partial_{\tau}\varphi_1-\varphi_1 \partial_{\tau}\varphi_2)=\\ \nonumber
&&-\frac{1}{2\pi i}\int_{-\infty}^{\infty}\mathrm{d}\tau
\Big[\log(s_2\tau^2+2m_2\tau+v_2){s_1\tau+m_1 \over s_1\tau^2+2m_1\tau+v_1}-\\ \nonumber
&&\log(s_1\tau^2+2m_1\tau+v_1){s_2\tau+m_2 \over s_2\tau^2+2m_2\tau+v_2}\Big]=\\ \nonumber
&&\log\left[{\frac{-s_1m_2+s_2m_1+i(s_2\sqrt{n_1^2-1}+s_1\sqrt{n_2^2-1})}
{s_1m_2-s_2m_1+i(s_2\sqrt{n_1^2-1}+s_1\sqrt{n_2^2-1})}}\right]\, .
\end{eqnarray}
Obviously this is  also a pure argument.
After cumbersome but straightforward calculation one can show that:
\be
\frac{(-s_1m_2+s_2m_1+i(s_2\sqrt{n_1^2-1}+s_1\sqrt{n_2^2-1}))}
{(s_1m_2-s_2m_1+i(s_2\sqrt{n_1^2-1}+s_1\sqrt{n_2^2-1}))}
{(\cosh\nu_2+i\cos\xi\sqrt{n_2^2-1}+n_2\sin\xi)\sinh\nu_1 \over (\cosh\nu_1-i\cos\xi\sqrt{n_1^2-1}-n_1\sin\xi)\sinh\nu_2}
=1\,
\ee
and therefore (\ref{purfase}) cancels the remaining
logarithmic terms in (\ref{purfaso}).
Collecting all we obtain:
\be
b^2S_0=2i\pi(N_1+N_2)-\log \lambda-2+\nu_2-\nu_1.
\ee

\section{Defect two-point function}
First let us  briefly explain how to derive the Cardy-Lewellen cluster condition
for defects in rational theories without multiplicities \cite{Petkova:2001ag,Sarkissian:2009aa}.
Suppose we have a two-dimensional rational conformal field theory with primary
fields $\Phi_i$. The vacuum state is attributed $i=0$.
A topological defect $X$ is a sum of projectors
\be\label{xpd}
X=\sum_{i}{\cal D}^{i}P^{i}
\ee
where
\be
P^{i}=\sum_{N,\bar{N}}(|i,N\rangle\otimes |i,\bar{N}\rangle)
(\langle i,N|\otimes \langle i,\bar{N}|)
\ee
Here $|i,N\rangle$ and ${|i,\bar{N}\rangle}$ are vectors of  orthonormal bases of left and right copies of the highest weight representations $R_i$
respectively.
Two-point functions with a defect $X$ insertion can be written as
\be\label{twopff}
\langle\Phi_{i}(z_1,\bar{z}_1)X\Phi_{i}(z_2,\bar{z}_2)\rangle={D^{i}\over
(z_1-z_2)^{2\Delta_i}(\bar{z}_1-\bar{z}_2)^{2\Delta_i}}\, ,
\ee

where
\be\label{conjik}
D^{i}={\cal D}^{i}C_{ii}
\ee
and $C_{ii}$ is a two-point function.

The fields $\Phi_{i}$ via the operator product expansion (OPE) form an algebra with structure constant $C_{ij}^{k}$ \cite{Belavin:1984vu,DiFrancesco:1997nk}:
\be\label{ope}
\Phi_{i}(z_1,\bar{z}_1)\Phi_{j}(z_2,\bar{z}_2)=\sum_{k}
{C_{ij}^{k} \over (z_1-z_2)^{\Delta_i+\Delta_j-\Delta_k}
(\bar{z}_1-\bar{z}_2)^{\Delta_{i}+\Delta_{j}-\Delta_{k}}}
\Phi_{k}(z_2,\bar{z}_2)+{\rm descendants}\, .
\ee
We need also to introduce the fusion number $N_{ij}^k$.
This is the number of occurrence of the field $\Phi_{k}$ in the operator product expansion
of $\Phi_{i}$ and $\Phi_{j}$. Here we assume that  $N_{ij}^k$ takes two values: $0$ and $1$.
Consider the following four-point correlation function with the defects insertions on a torus:
\be\label{fourp}
\langle\Phi_{j}(z_1,\bar{z}_1)\Phi_{i}(z_2,\bar{z}_2)X\Phi_{i}(z_3,\bar{z}_3)\Phi_{j}(z_4,\bar{z}_4)X\rangle\, .
\ee
Using (\ref{ope}) and (\ref{twopff}) one can compute (\ref{fourp}) in two pictures.
In the first picture at the beginning we use OPE (\ref{ope}) for the pairs $\Phi_{j}(z_1,\bar{z}_1)\Phi_{i}(z_2,\bar{z}_2)$
and $\Phi_{i}(z_3,\bar{z}_3)\Phi_{j}(z_4,\bar{z}_4)$ and then (\ref{twopff}) for the fields produced in this process.
This results in

\be\label{frprd}
\sum_k D^{k}D^0
\left(C_{ij}^{k}{\cal F}_{k}\left[\begin{array}{cc}
i&i\\
j&j \end{array}\right]\right)^2\, ,
\ee
where ${\cal F}_{k}\left[\begin{array}{cc}
i&i\\
j&j \end{array}\right]$ is the so called conformal block \cite{Belavin:1984vu,DiFrancesco:1997nk} giving the contribution of
the descendant fields in the OPE (\ref{ope}). It appears squared since it is separately produced by the left and right modes.

In the second picture we move the field $\Phi_{j}(z_1,\bar{z}_1)$ to the rightmost position:
\bea\label{secprd}
\langle\Phi_{i}(z_2,\bar{z}_2)X\Phi_{i}(z_3,\bar{z}_3)\Phi_{j}(z_4,\bar{z}_4)X\Phi_{j}(z_1,\bar{z}_1)\rangle
\eea
and then use twice (\ref{twopff}) resulting in
\be
D^{i}D^{j}
\left({\cal F}_{0}\left[\begin{array}{cc}
i&j\\
i&j \end{array}\right]\right)^2
+\cdots\, .
\ee

Using the fusing matrix:
\bea
{\cal F}_{k}\left[\begin{array}{cc}
i&i\\
j&j \end{array}\right]
=
\sum_m F_{km}\left[\begin{array}{cc}
j&j\\
i&i\end{array}\right] {\cal F}_{m}\left[\begin{array}{cc}
i&j\\
i&j \end{array}\right]\, ,
\eea
we obtain
\bea\label{defalg}
\sum_k D^0D^{k}\left(C_{ij}^{k}
 F_{k0}\left[\begin{array}{cc}
j&j\\
i&i\end{array}\right]\right)^2=
D^{i}D^{j}\, .\hspace{1cm}
\eea
This is the Cardy-Lewellen cluster condition for defects.

Using that for rational conformal field theory the structure constants and the fusion matrix satisfy the relation \cite{Behrend:1999bn}
\be\label{cpij}
C^p_{ij}F_{p,0}\left[\begin{array}{cc}
j&j\\
i&i \end{array}\right]={\xi_i\xi_j\over \xi_0\xi_p}\, ,
\ee
where
\be
\xi_i=\sqrt{C_{ii}F_i}\, ,
\ee
and
\be
F_i\equiv F_{0,0}\left[\begin{array}{cc}
i&i\\
i&i \end{array}\right]\, ,
\ee

the Cardy-Lewellen condition for defects
(\ref{defalg}) simplifies to

\be\label{ddok}
\sum_k D^0D^{k} N_{ij}^k\left({\xi_i\xi_j\over \xi_0\xi_k}\right)^2=D^{i}D^{j}\,.
\ee
Define $\Psi^k$ as
\be\label{dkdo}
{D^{k}\over D^0}=\Psi^k\left({\xi^k\over \xi^0}\right)^2\, .
\ee
Eq. (\ref{ddok}) becomes the following equation for $\Psi^k$
\be\label{nbk}
\sum_k \Psi^k N_{ij}^k=\Psi^i\Psi^j\, .
\ee
And finally to find the coefficient ${\cal D}^{i}$ of the defects expansion to projectors we should, according
to (\ref{conjik}),  divide $D^{i}$ by the two-point function.

Let us now apply this machinery to the Liouville theory.
Liouville theory is a non-rational theory, but we can overcome the difficulties caused by the infinite number of primaries.
First of all
it is shown in \cite{Sarkissian:2011tr} that the relation (\ref{cpij}) works also in
diagonal non-rational theories. In particular it is shown in \cite{Sarkissian:2011tr} that in the Liouville theory
\be\label{xil}
\xi(\alpha)={\sqrt{W(0)W(Q)}\over W(\alpha)}\, .
\ee
and (\ref{cpij}) takes the form:
\be\label{cfal}
C_{\alpha_1,\alpha_2}^{\alpha_3}F_{\alpha_3,0}\left[\begin{array}{cc}
\alpha_1&\alpha_1\\
\alpha_2 &\alpha_2 \end{array}\right]=W(0){W(\alpha_3)\over W(\alpha_1)W(\alpha_2)}\, ,
\ee
where $W(\alpha)$ is ZZ function (\ref{zzfu}).
The second problem is that in the Liouville theory the OPE of primary fields with generic $\alpha_1$, and $\alpha_2$
contains infinite number of intermediate primary states, which makes the use of the equation (\ref{nbk})
rather problematic. This difficulty can be resolved via Teschner's trick \cite{Teschner:1995yf}. Teschner's tricks relies on the existence
of  degenerate fields in the Liouville field theory. The fields $V_{\alpha}$ with $\alpha$ belonging to the set
\be\label{deger}
\alpha_{m,n}={1-m\over 2b}+{1-n\over 2}b\, ,\hspace{1cm} m,n\in \mathbb{N}
\ee
produce in the OPE with other fields just a finite number of the fields. Teschner's trick suggests to
take as $\Phi_{j}$ one of the fields $V_{\alpha_{m,n}}$. This choice will yield only finite number of terms in the l.h.s.
of (\ref{nbk}). The simplest of the fields (\ref{deger}) is $V_{-b/2}$. With a generic field $V_{\alpha}$ it has the OPE:
\be
V_{\alpha}V_{-b/2}\sim C_{-b/2,\alpha}^{\alpha-b/2}V_{\alpha-b/2}+C_{-b/2,\alpha}^{\alpha+b/2}V_{\alpha+b/2}\,.
\ee

With $j=-{b\over 2}$,  $i=\alpha$, and $k=\alpha \pm b/2$, the equations (\ref{nbk}) and (\ref{dkdo}) take the form:
\be\label{psik}
\Psi(\alpha)\Psi(-b/2)=\Psi(\alpha-b/2)+\Psi(\alpha+b/2)\, ,
\ee
and
\be\label{dkdol}
{D(\alpha)\over D(0)}=\Psi(\alpha)\left({W(0)\over W(\alpha)}\right)^2
\ee
The solution of the equation (\ref{psik}) is
\be
\Psi_{m,n}(\alpha)={\sin(\pi m b^{-1}(2\alpha- Q))\sin(\pi nb (2\alpha-Q))\over \sin(\pi mb^{-1} Q)\sin(\pi nb Q)}\, ,
\ee
Using (\ref{dkdol}) we obtain for the defect two-point function

\be\label{discrett}
D_{m,n}(\alpha)=-{2\sqrt{2}\sin(\pi m b^{-1}(2\alpha- Q))\sin(\pi nb (2\alpha-Q))\over W^2(\alpha)}\, .
\ee
And finally dividing on $S(\alpha)$ (\ref{reflal}) we get
\be\label{discret}
{\cal D}_{m,n}(\alpha)={\sin(\pi m b^{-1}(2\alpha- Q))\sin(\pi nb (2\alpha-Q))\over \sin\pi b^{-1}(2\alpha-Q)\sin\pi b(2\alpha-Q)}\, .
\ee
Note that the defect given by $(m,n)=(1,1)$ is the identity defect.

But this is not the end of the story.
Let us now explain how to obtain two-point function for the continuous family of defects.
We will use the strategy developed in \cite{Fateev:2000ik,Fateev:2010za} in the context of the Liouville and Toda theories
with a boundary.
Assume that we have a family of defects parameterized by $\kappa$.
In this case  $D(-b/2)$, which is the
two-point function of the degenerate field $V_{-b/2}$ in the presence of defect, will be a function
of $\kappa$ and $b$. Denote the ratio $D(-b/2)/D(0)$ by $A(\kappa,b)$ and define
\be\label{dtilp}
D(\alpha)={\tilde{\Psi}(\alpha)\over W^2(\alpha)}\, .
\ee
Substituting $A(\kappa,b)$ and $\tilde{\Psi}(\alpha)$ in (\ref{ddok}) again for $j=-{b\over 2}$,  $i=\alpha$, and $k=\alpha \pm b/2$, we obtain a linear equation for $\tilde{\Psi}(\alpha)$:

\be\label{tsik}
\left({W(-b/2)\over W(0)}\right)^2A\tilde{\Psi}(\alpha)=\tilde{\Psi}(\alpha-b/2)+\tilde{\Psi}(\alpha+b/2)\, .
\ee

The solution of (\ref{tsik}) is indeed a one-parametric family,
\be\label{funcla}
\tilde{\Psi}_s(\alpha)=-2^{1/2}\cosh(2\pi s(2\alpha-Q))\, ,
\ee
 with a parameter $s$ related to $A$ by
\be\label{rela}
2\cosh 2\pi bs=A\left({W(-b/2)\over W(0)}\right)^2\, .
\ee

Substituting (\ref{funcla}) in (\ref{dtilp}) we obtain for $ D_{s}(\alpha)$ and ${\cal D}_{s}(\alpha)$ respectively
\be\label{contt}
D_{s}(\alpha)=-{2^{1/2}\cosh(2\pi s(2\alpha-Q))\over W^2(\alpha)}\, .
\ee
\be\label{cont}
{\cal D}_{s}(\alpha)={\cosh(2\pi s(2\alpha-Q))\over 2\sin\pi b^{-1}(2\alpha-Q)\sin\pi b(2\alpha-Q)}\, .
\ee
We would like to finish by a remark on the world-volume of the defects  (\ref{discret}) and (\ref{cont}).
Recall the notion of the defect world-volume \cite{Fuchs:2007fw}.

The values of the Liouville fields $\phi_1$ and  $\phi_2$ on a point $\tau$ of the defect line
form a point $(\phi_1(\tau),\phi_2(\tau))$ in the plane $\mathbb{R}^2$. The set of all such points
 may be restricted to belong to a submanifold $Q$ of the plane $\mathbb{R}^2$, depending on the defect condition. The submanifold $Q$ is called the world-volume of the defect. It can be shown that the world-volume
of the defects (\ref{cont}) coincide with all $\mathbb{R}^2$, which means that there are no constraints
on the values of the fields $\phi_1$ and $\phi_2$. But the world-volume of the defects (\ref{discret}) is a
one-dimensional. It can be easily seen for the identity defect ${\cal D}_{1,1}$, since for the identity defect
there is no discontinuity in the value of Liouville field and therefore $\phi_1$ and $\phi_2$ satisfy $\phi_1(\tau)=\phi_2(\tau)$ in any point
$\tau$ of the defect line.

\end{document}